\documentclass[10pt,aps,prb,twocolumn,amsmath,amssymb,floatfix]{revtex4-2}

\usepackage[dvipsnames]{xcolor}
\usepackage{graphicx}
\usepackage[colorlinks,allcolors=blue,plainpages=false,pdfpagelabels]{hyperref}
\usepackage{txfonts}
\usepackage{amsmath,bm}
\usepackage{tikz}
\usetikzlibrary{positioning, shapes}
\pgfdeclarelayer{background}
\pgfdeclarelayer{foreground}
\pgfsetlayers{background,main,foreground}

\usetikzlibrary{shapes.misc}
\tikzset{cross/.style={cross out, draw=black, minimum size=2*(#1-\pgflinewidth), inner sep=0pt, outer sep=0pt},cross/.default={1pt}}

\usepackage[normalem]{ulem}

\newcommand{\bra}[1]{\ensuremath{\langle{#1}|\,}}
\newcommand{\ket}[1]{\ensuremath{\,|{#1}\rangle}}
\newcommand{\braket}[1]{\ensuremath{\langle{#1}\rangle}}

\renewcommand{\Re}{\operatorname{Re}}

\begin{document}
\title{Braiding of Majorana bound states in a driven-dissipative Majorana box setup}
	
\author{Kunmin Wu}

\author{Sadeq S. Kadijani}

\author{Thomas L. Schmidt}
\email{thomas.schmidt@uni.lu}
\affiliation{Department of Physics and Materials Science, University of Luxembourg, 1511 Luxembourg, Luxembourg}

\date{\today}
	
\begin{abstract}
    We investigate a system of Majorana box qubits, where each of the Coulomb blockaded boxes is driven by an applied AC voltage and is embedded in a dissipative environment. The AC voltage is applied between a pair of quantum dots, each of which is coupled by tunneling to a Majorana box qubit. Moreover, the dissipation is created by the coupling to an electromagnetic environment. Recent work has shown that in this case the Majorana bound states which form the computational basis can emerge as dark states, which are stabilized by the dissipation. In our work, we show that the same platform can be used to enable topological braiding of these dissipative Majorana bound states. We show that coupling three such Majorana boxes allows a braiding transformation by changing the tunnel amplitudes adiabatically in time.
\end{abstract}
	
\maketitle

\section{Introduction} \label{sec:int}

Majorana bound states (MBSs) have been proposed to exist as zero-energy quasiparticles in several solid-state systems. Among the most promising experimental platforms are semiconductor nanowires with Rashba spin-orbit coupling in a magnetic field and subject to a superconducting proximity effect \cite{oreg2010helical, sato2009topological, lutchyn2010majorana, lutchyn2018majorana, mourik2012signatures, rokhinson2012fractional, das2012zero, deng2012anomalous, churchill2013superconductor, finck2013anomalous}, chains of ferromagnetic atoms on top of a superconductor \cite{choy2011majorana, kjaergaard2012majorana, martin2012majorana, nadj2013proposal, pientka2013topological, klinovaja2013topological, braunecker2013interplay, vazifeh2013self, nadj2014observation, kim2014helical, poyhonen2014majorana, peng2015strong, ruby2015end, pawlak2016probing, feldman2017high, jeon2017distinguishing, kim2018toward} or quantum spin Hall edge states coupled to superconductors \cite{read2000paired, fu2008superconducting, fu2009josephson, hart2014induced, pribiag2015edge}. The vibrant activity in this field is driven to a large extent by the intriguing properties of such Majorana bound states, in particular their non-Abelian exchange statistics \cite{moore1991nonabelions, kitaev2006anyons, nayak2008non, hasan2010colloquium, alicea2012new, beenakker2013search, sarma2015majorana, aguado2017majorana}.

This nontrivial exchange statistics was first discussed in the context of two-dimensional $p$-wave superconductors, where vortices of the superconducting order parameter host MBSs \cite{ivanov2001non, stern2013topological, stanescu2013majorana}. Two such MBSs span a two-dimensional low-energy Hilbert space, and a spatial adiabatic exchange (``braiding'') of the positions of two MBSs acts as a nontrivial unitary transformation on the system state within this ground state subspace. In the experimental realizations mentioned above, MBS typically emerge at the ends of one-dimensional systems \cite{kitaev2001unpaired}, but even in those cases adiabatic exchanges can often be implemented using, e.g., T-junctions \cite{alicea2011non} or changing superconducting fluxes \cite{ivanov2001non, van2012coulomb}.

The non-Abelian statistics of MBSs is also the cornerstone of their proposed application in topological quantum computing \cite{freedman2003topological, nayak2008non, wang2010topological, aasen2016milestones}. In that case, the Hilbert space spanned by the MBSs can be used to define a qubit state, and information in such a state would be stored nonlocally, rendering it robust to certain perturbations. Braiding them would allow the implementation of certain qubit gates. 

So far, MBSs have mainly been considered in closed quantum systems. However, it was shown early on that open quantum systems can give rise to MBS as well \cite{plenio2002entangled, Zoller2011, Bardyn2013}. An open quantum system is embedded in an environment and in the simplest case its reduced density matrix evolves according to the Lindblad equation \cite{lindblad1976generators, lindblad2001non, breuer2002theory, weiss2012quantum}. While most quantum states decay in the presence of damping, certain dark states \cite{gardiner2004quantum} of the Hilbert space may be immune to such dissipation \cite{santos2020possible, iemini2015localized, iemini2016dissipative}. In this case, damping can stabilize a dark state subspace which can contain MBSs \cite{touzard2018coherent, geerlings2013demonstrating, lu2017universal}. This idea is behind the proposed realization of a Kitaev chain \cite{kitaev2001unpaired} in dissipative cold-atom systems \cite{Zoller2011}.

The topological protection of MBSs rests on a conserved fermionic parity in the system. In that respect, MBSs emerging in quantum dots (QDs) \cite{deng2016majorana} in the Coulomb blockade regime are beneficial because a large charging energy can protect them from charge fluctuations. Recently, it was shown that an open-system scenario can be created as well for MBSs in Majorana boxes coupled to tunneling junctions \cite{Gau2020, gau2020driven}. In this case, the Lindblad dynamics can in fact stabilize certain one-qubit and two-qubit Majorana states which are interesting for quantum computation \cite{verstraete2009quantum, fujii2014measurement}. However, in those works, it was not discussed how braiding could be performed in such MBSs.

In our work, we will consider a setup consisting of Majorana boxes placed in a dissipative environment. To engineer dissipation, the Majorana boxes are linked via tunnel junctions to a driven fermionic reservoir and damping is created by an electromagnetic Ohmic environment which affects tunneling \cite{breuer2002theory, weiss2012quantum}. Depending on the chosen system parameters, this system can stabilize a dark state subspace of MBSs. We extend this setup to a network of three Majorana boxes. In this case, we show that an adiabatic periodic change of the system parameters can be used to drive the dark states and to implement a braiding operation on the MBS, which is manifested in a quantized topological winding number. 

The structure of this article is as follows: in Sec.~\ref{sec:the_open_system_with_Majorana boxe}, we briefly revisit the main results of Ref.~\cite{Gau2020} and show how a driven-dissipative Majorana box system can give rise to a dark state subspace. Next, in Sec.~\ref{sec:stabilization_in_Majorana boxe}, we start from the general Lindblad dynamics of a single Majorana box system and show how its dark states can be driven by changing the parameters, giving rise to quantized winding numbers. In Sec.~\ref{sec:braiding_in_DFS} we show how to implement such a braiding transformation in a system consisting of three Majorana boxes. Finally, we present our conclusions in Sec.~\ref{sec:conclusion_and_outlooks}.

\section{An open system with Majorana boxes} \label{sec:the_open_system_with_Majorana boxe}

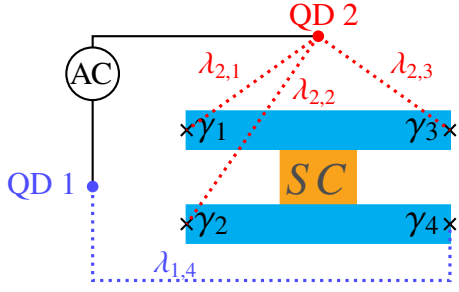
\begin{figure}[t]
    \begin{tikzpicture}
    \filldraw[cyan, thick] (2.25,0.5) rectangle (5.75,1);
    \draw[black] (2.25, 0.75) node[anchor=west]{\Large $\gamma_{1}$};
    \draw[black, thick] (2.25, 0.75) node[cross=3pt, rotate=90]{};
    \draw[black] (5.05, 0.75) node[anchor=west]{\Large $\gamma_{3}$};
    \draw[black, thick] (5.75, 0.75) node[cross=3pt, rotate=90]{};

    \filldraw[cyan, thick] (2.25,-0.75) rectangle (5.75,-0.25);
    \draw[black] (2.25, -0.5) node[anchor=west]{\Large $\gamma_{2}$};
    \draw[black, thick] (2.25, -0.5) node[cross=3pt, rotate=90]{};
    \draw[black] (5.05, -0.5) node[anchor=west]{\Large $\gamma_{4}$};
    \draw[black, thick] (5.75, -0.5) node[cross=3pt, rotate=90]{};

    \filldraw [red] (4,2) circle (2pt); 
    \draw[red] (3.5, 2.25) node[anchor=west]{\large QD $2$};
    \draw[dotted, red, very thick] (4, 2) -- (2.25, 0.75) node[red] at (2.7, 1.6) { \large $\lambda_{2,1}$};
    \draw[dotted, red, very thick] (4, 2) -- (5.75, 0.75) node[red] at (5.25, 1.6) { \large $\lambda_{2,3}$};
    \draw[dotted, red, very thick] (4, 2) -- (2.25, -0.5) node[red] at (3.95, 1.25) { \large $\lambda_{2,2}$};

    \filldraw [blue!70] (1,0) circle (2pt);
    \draw[blue!70] (-0.25, 0) node[anchor=west]{\large QD $1$};
    \draw[dotted, blue!70, very thick] (1, 0) -- (1, -1.25) -- (5.75, -1.25)-- (5.75, -0.5) node[blue!70] at (2.1, -1.05) { \large $\lambda_{1,4}$};

    \filldraw [color=black, fill=white, thick] (1,1.5) circle (0.35) node[anchor=west] at (0.575, 1.5) {\large AC};

    \begin{pgfonlayer}{background}
    \filldraw[YellowOrange, thick] (3.5,-0.25) rectangle (4.5,0.5);
    \draw[black!70] (3.455, 0.09) node[anchor=west]{\LARGE $SC$};
    \draw [black, thick] (4,2)--(1,2)--(1,0);
    \end{pgfonlayer}

    \end{tikzpicture}
    \caption{Two nanowires (blue rectangles) are connected by a superconducting (SC) bridge (orange rectangle). The four edge modes of these nanowires are Majorana bound states (MBSs) marked as $\gamma_{\nu}$. These MBSs are tunnel coupled with the quantum dots QD $1$ and QD $2$ shown by the dashed line, and $\lambda_{j,\nu}$ denotes the tunnel amplitude from QD $d_{j}^{\dagger}$ to MBS $\nu$. An AC voltage is applied between two QDs and drives the system.}
    \label{fig:2by2_setup01}
\end{figure}

The basic building block of the system under study is shown in Fig.~\ref{fig:2by2_setup01}. A single Majorana box qubit consists of two Majorana nanowires coupled to a common, floating superconductor. Its small size leads to a significant charging energy for the Majorana system. Moreover, each of the Majorana bound states is coupled via tunneling to one of two quantum dots QD1 and QD2. An AC voltage between the two quantum dots can be used to pump electrons between them. Moreover, we assume that the whole system is embedded in an electromagnetic environment, which mainly affects the phases of the tunneling amplitudes between the quantum dots and the Majorana bound states. The Hamiltonian and the approximations we use are described in detail in Ref.~\cite{Gau2020}, but we reiterate the essential steps to make our discussion self-contained.

We consider the resulting system as an open quantum system, where the electromagnetic field modes constitute a bosonic environment for the Majorana box. The system Hamiltonian consists of the Hamiltonian of the Majorana box ($H_{\rm box}$) and that of the quantum dots ($H_{\rm QD}$), as well as Hamiltonians for the (time-dependent) driving field between the QDs $H_{\rm drive}(t)$, the tunneling between the QDs and MBSs $H_{\rm tun}$, and the bosonic environment $H_{\rm env}$. The total Hamiltonian therefore reads
\begin{align}
	H(t) &= H_{\rm box} + H_{\rm QD} + H_{\rm tun} + H_{\rm env} +  H_{\rm drive}(t) . \label{Hamil:the_entire_system}
\end{align}
At energies below the superconducting gap, the Majorana bound states $\gamma_{\nu}$ ($\nu \in \{1,\ldots,4\}$) and the Cooper pairs in the superconducting island are the only degrees of freedom of the Majorana box. The four MBSs span a four-dimensional Hilbert space. The Hamiltonian of the Majorana box is determined by its charging energy $E_{C}$ and its total electron number $\hat{N}$, which comprises the Cooper pairs as well as the electrons in the MBSs, 
\begin{align}
        H_{\rm box} = E_{C} \left( \hat{N} - N_{g} \right)^2. \label{Hamil:Majorana_box}
\end{align}
We assumed that the direct overlaps between different MBSs are negligible. The parameter $N_{g}$ accounts for the presence of a back-gate voltage which can be used to offset the charging energy and thus control the ground state of the box. When $N_{g}=1$ the ground states have the fermion number $\hat{N} = 1$, and this condition is fulfilled for the two degenerate Majorana states with odd parity. The Majorana box will be trapped in this ground state subspace as long as the energy scale of any dynamics is lower than the charging energy $E_{C}$ and the superconducting gap \cite{Gau2020}. In the following, we will assume this to be the case, so we can drop the term $H_{\rm box}$ because it will merely act as a constant term in this reduced ground-state subspace.

The Majorana box is connected by tunneling to two quantum dots. We assume that each QD can be described by a single fermionic level, 
\begin{align}
    H_{\rm QD} = \sum_{j=1}^2 \epsilon_j d_j^\dag d_j    
\end{align}
with fermionic annihilation operators $d_j$. Each of the two QDs is coupled to one or more MBSs by electron tunneling,
\begin{align}
    H_{\rm tun} = t_{0} e^{-i\hat{\phi}} \sum_{jv} e^{i \delta_{j} } \lambda_{j \nu} e^{ - i \beta_{j \nu} } d_{j}^{\dagger} \gamma_{\nu} + \text{h.c.} ,   \label{Hamil:tunneling_Hamiltonian}  
\end{align}
where $t_{0}$ is the overall tunnel amplitude between QDs and the Majorana box. Moreover, $\hat{\phi}$ is a phase operator which obeys the commutator $[ \hat{\phi}, \hat{N} ] = -i$ and which is necessary to account for the change of electron number in the Majorana box upon tunneling. Moreover, the (dimensionless) tunable tunnel amplitudes between the QD electron $d_{j}$ and MBS $\gamma_{\nu}$ are complex functions with real amplitudes $\lambda_{jv}$ and phases $\beta_{jv}$ in the interval $\left[ 0, \; \pi \right)$. Since the total electron number is fixed, we assume that the two quantum dots contain in total a single electron in order to allow for electron tunneling between the quantum dots.

The bath Hamiltonian $H_{\rm env}$ is a collection of harmonic oscillators $b_m$ which constitute an Ohmic bath. Their main effect of the bath is to change the phases of the tunneling amplitudes between the QD $j$ and the MBSs,
\begin{align}
	\delta_{j} &= \sum_{m} g_{jm}  \left( b_{m} + b_{m}^{\dagger}     \right),  \label{eq:EM_fluctuation_at_QDj}
\end{align}
with real-valued constants $g_{jm}$.

\subsection{Dynamics in the low-energy sector} \label{subsec:dissipation_in_Hamiltonian}

Assuming tunneling to be weak, we can eliminate the tunneling term up to the second order in $t_0$ by using a Schrieffer-Wolff (SW) transformation. Before the SW transformation, the Hamiltonian can be written as a sum of an unperturbed term $H_{0} = H_{\rm QD} + H_{\rm env} + H_{\rm drive}(t)$ and a small perturbation $H_{\rm tun}$. Then, the SW transformation is implemented using a unitary transformation $e^{S}$ such that $H' = e^{S} (H_{0} + H_{\rm tun}) e^{-S}$, where the generator $S$ is of the order of the tunneling amplitude. By choosing $S$ such that $H_{\rm tun} + [S, H_{0}] = 0$, the SW transformation yields a transformed Hamiltonian $H_{\rm eff} \approx H_{0} + [S, H_{\rm tun}]/2 = H_0 + H_{\rm cot}$, where terms of order $t_0$ have been eliminated. After the SW transformation, we thus obtain an effective cotunneling contribution,
\begin{align}
    H_{\rm cot} &=  V_{\rm cot} + V_{0}, \label{Hamil:cotunneling_single_box} \\
    V_{\rm cot} &= 2 g_{0} \left( e^{ i \delta_{12} } A_{12} d_{2}^{\dagger} d_{1} + \text{h.c.} \right), \nonumber \\
    V_{0} &=  g_{0} \sum_{j=1,2} A_{jj} d_{j}^{\dagger} d_{j}, \nonumber
\end{align}
where $g_{0} \equiv t_{0}^2 / E_{C}$ and $\delta_{12} = \delta_{2} - \delta_{1}$ with $\delta_{j}$ defined in Eq.~\eqref{eq:EM_fluctuation_at_QDj}. The operators $A_{jk}$ account for the tunneling trajectories of the electrons through the Majorana boxes,
\begin{align}
    A_{jk} = \sum_{ \mu < \nu } \left[ \lambda_{j \nu} \lambda_{k\mu} e^{ i \left( \beta_{k \mu} - \beta_{j \nu} \right)} - \lambda_{j \mu} \lambda_{k \nu} e^{ i \left( \beta_{k \nu} - \beta_{j \mu} \right)} \right] \gamma_{\mu} \gamma_{\nu}. \label{eq:operator_A}
\end{align}

Finally, we discuss the effect of the time-dependent drive. It is provided by an AC voltage with frequency $\omega_{0}$ between the pair of quantum dots QD1 and QD2,
\begin{align}
    H_{\rm drive}(t) = 2 \mathcal{A}  \cos(\omega_0 t) d_1^\dag d_2 + \text{h.c.}
\end{align}
When the AC frequency is close to the energy difference between two QDs, $\omega_0 \approx \epsilon_{2} - \epsilon_{1}$, tunneling becomes resonant
and it is possible to use the rotating wave approximation (RWA) \cite{breuer2002theory, gardiner2004quantum}. The RWA corresponds to going to a rotating frame and neglecting the fast oscillating terms proportional to $e^{ \pm 2 i \omega_0 t }$. In this case the drive term takes the form,
\begin{align}
    H_{\rm drive}^{\rm RWA} = \mathcal{A} d_{1}^{\dagger} d_{2} + \text{h.c.} .
\end{align}
As a result, the total effective Hamiltonian now describes the Majorana bound states and the quantum dot electrons, which are coupled to a bosonic bath via the term $V_{\rm cot}(t)$ in the cotunneling Hamiltonian \eqref{Hamil:cotunneling_single_box}, 
\begin{align}
	H_{\rm eff} &= H_{\rm QD} + H_{\rm cot} + H_{\rm env} + H_{\rm drive}^{\rm RWA}. \label{Hamil:effective_low_energy_theory}
\end{align}
This effective Hamiltonian is the starting point for describing the dynamics of the reduced system density matrix using a Lindblad master equation.

\subsection{The Lindblad master equation for the Majorana bound states} \label{subsec:Lindblad_Eq_single_MB}

In the low-energy limit and using the other approximations explained in Sec.~\ref{subsec:dissipation_in_Hamiltonian}, the coupling between the system and the environment is described by $H_{\rm cot}$, see Eq.~\eqref{Hamil:cotunneling_single_box}. Next, we will use the Born-Markov approximation, in which we assume that the coupling to the bath is weak and that the bath relaxes fast to equilibrium. This approximation will give rise to a Master equation for the MBS system and QDs which will be of Lindblad form. The Lindblad equation maps the reduced system density matrix from its initial state to the steady state while ensuring that the basic properties of the density matrix remain fulfilled. 

The reduced system state becomes factorized on a timescale that corresponds to the inverse of the dissipative gap of the reduced Lindbladian governing exclusively the Majorana sector. Hence, it is reasonable to approximate $\rho(t \to \infty)$ as $\rho_{M}(t \to \infty) \otimes \rho_{QD}$, where $\rho_{M}$ and $\rho_{QD}$ represent the reduced density matrices of the MBSs and QDs, respectively. After tracing over the environment and QD degrees of freedom, one obtains the Lindblad equation for the Majorana box \cite{Gau2020},
\begin{align}
   \frac{ d }{ d t } \rho_{M} &= \mathcal{L} \rho_{M}(t) \nonumber \\
   &\approx \sum_{ j \neq k = 1}^2 \Gamma_{jk} \left[ A_{jk} \rho_{M}(t) A_{jk}^{\dagger} - \frac{1}{2} \left\{ A_{jk}^{\dagger} A_{jk}, \rho_{M}(t) \right\}\right], \label{eq:Lindblad_equation_2QDs}
\end{align}
where $A_{jk}$ is the jump operator defined in Eq.~\eqref{eq:operator_A}, and $\Gamma_{jk}$ is the corresponding transition rate, which is given by
\begin{align}
    \Gamma_{21} &= 2 \Re \left[ \Lambda_{-} \right], \; \Gamma_{12} = 2 \Re \left[ \Lambda_{+} \right], \label{eq:transition_rates_Gamma_1221} \\
    \Lambda_{\pm} &= 4 g_{0}^{2} \int_{0}^{\infty} ds \; e^{\pm i \omega_{0} s} \;  \braket{ e^{ \pm i  \left[ \delta_{12}(s) - \delta_{12}(0) \right] } }_{\rm env}  \label{eq:the_bath_part_in_RF_equation} \notag \\
    &= 4 g_{0}^{2} \int_{0}^{\infty} ds \; e^{\pm i \omega_{0} s} \; e^{ J_{\rm env}(s) }, 
\end{align}
where the correlation function of the bosonic bath is
\begin{align}
    J_{\rm env}(t) &= \int \frac{d \omega}{ \pi } \frac{ \mathcal{J}(\omega) }{ \omega^2 } \nonumber \\
    & \times \left[ \coth \left( \frac{ \omega }{ 2T } \right) \; \left( \cos \left( \omega t \right) - 1 \right) - i \sin \left( \omega t \right) \right]. \label{eq:correlation_function}
\end{align}
We assume the bath to be Ohmic, so the spectral density $\mathcal{J}(\omega)$ is proportional to the bath frequency $\omega$ for frequencies up to a cut-off frequency $\omega_{c}$ \cite{weiss2012quantum}. 

From the bath correlation function \eqref{eq:correlation_function}, one can prove that $J_{\rm env}(-t-i/T) = J_{\rm env}(t)$, which leads to $\Lambda_{-} = e^{- \omega_{0}/T } \Lambda_{+}$. Therefore, the forward and backward transition rates are related as,
\begin{align}
    \Gamma_{21} = e^{ -\omega_{0}/T } \Gamma_{12}, \label{eq:two_transition_rates_relation}
\end{align}
leading to a suppression of $\Gamma_{21}$ with respect to $\Gamma_{12}$.

In the Lindblad equation~\eqref{eq:Lindblad_equation_2QDs}, we neglected the unitary time evolution $-i [H_{LS}, \rho_{M}]$ with the ``Lamb shift'' Hamiltonian $H_{LS} = \sum_{j,k} h_{jk} A_{jk}^{\dagger} A_{jk}$. The prefactor $h_{jk}$ is the imaginary part of $\Lambda_{\pm}$ \eqref{eq:the_bath_part_in_RF_equation}. As the steady state is a mixed state in the basis of dark states, each is which is annihilated by $A_{jk}$, the unitary part vanishes when the system reaches the steady state. Hence, the unitary dynamics do not influence the dark-state subspace where braiding will be implemented.

So far, we have reviewed the dissipative dynamics of the Majorana box within the Born-Markov approximation. The ensuing discussion rests on the fact that the steady-state density matrices which obey $\mathcal{L}\rho_{M}(t \to \infty) = 0$ form a dark state subspace, which allows the stabilization of certain Majorana qubit states.

\section{Topological phases in open Majorana boxes} \label{sec:stabilization_in_Majorana boxe}

In this section, we consider a general expression for the steady state density matrix in the open system consisting of a single Majorana box coupled to two QDs. The dissipation in this system is described by the Lindblad equation \eqref{eq:Lindblad_equation_2QDs} when the correlation time of the system is much longer than that of the bath. Then, we follow Ref.~\cite{Zoller2011} to show in which cases there is topological order in the steady state. In the end, we provide a stabilization protocol which drives the system in Fig.~\ref{fig:2by2_setup01} to a pure state with topological order.

\subsection{General form of steady-state density matrix} \label{subsec:2by2_DM_SS_general_form}

We will first explain how a topological transformation can be implemented within the dark state subspace generated by the steady-state solutions of the Lindblad equation~\eqref{eq:Lindblad_equation_2QDs}. That equation holds for a single Majorana box and the associated two-dimensional dark state subspace for a given fermionic parity. Non-Abelian braiding is not possible in a two-dimensional Hilbert space because it is always possible to find a basis which makes the braid matrix diagonal and thus leads at most to (Abelian) exchange phases. Nevertheless, we present the ensuing discussion to explain in a compact form how braiding will be implemented later in a larger Majorana box qubit system. 

To keep this discussion general, we consider a Lindblad equation of the form~\eqref{eq:Lindblad_equation_2QDs} for a density matrix $\rho(t)$ but with an arbitrary $2 \times 2$ jump operator $K$ and a general damping amplitude $\Gamma$,
\begin{align}
    \frac{ d }{ d t } \rho(t) &= \hat{\mathcal{L}} \left[ K \right] \rho(t) =
    \Gamma \left[ K \rho(t) K^{\dagger} - \frac{1}{2} \left\{ K^{\dagger} K, \rho(t) \right\}\right].
    \label{eq:Lindblad_equation_gen}
\end{align} 
From the relation between the two transition rates in Eq.~\eqref{eq:two_transition_rates_relation}, one finds that the rate $\Gamma_{21}$ is exponentially suppressed for $T \ll \omega_{0}$ in the low-energy regime. Thus, we only consider the trajectories from QD $1$ to QD $2$. Then, the jump operator and the density matrix can be parametrized as follows,
\begin{align}
    K &= \bm{k} \cdot \bm{\sigma} \label{eq:2by2_general_K} \\
    \rho(t) &= \frac{1}{2} \left( \sigma_{0} + \bm{n}(t) \cdot \bm{\sigma}   \right)  \label{DM:2by2_general_form_Bloch_vector}
\end{align}
with $\bm{k} = ( k_{x} e^{ i \phi_{x} }, \; k_{y} e^{ i \phi_{y} }, \; k_{z} e^{ i \phi_{z} } )$ and $\bm{n} = ( n_{x}, \; n_{y}, \; n_{z} )$, where $k_{x,y,z} \in \mathbb{R}$, $n_{x,y,z} \in \mathbb{R}$ and $\phi_{x,y,z} \in [0, \; \pi)$. Moreover, $\sigma_{0}$ denotes the identity matrix and $\bm{\sigma}$ is the vector of Pauli matrices. As the product of two Pauli matrices satisfies $\sigma_{j}\sigma_{k} = \delta_{jk} \sigma_{0} + i \epsilon_{jkl} \sigma_{l}$ with the Levi-Civita symbol $\epsilon_{jkl}$ one finds $\left( \bm{\nu}_1 \cdot  \bm{\sigma} \right) \left( \bm{\nu}_2 \cdot  \bm{\sigma} \right) = \left( \bm{\nu}_1 \cdot \bm{\nu}_2 \right) \sigma_0 + i \bm{\sigma} \cdot \left( \bm{\nu}_1 \times \bm{\nu}_2  \right)$ for two arbitrary vectors $\bm{\nu}_1$ and $\bm{\nu}_2$. This makes it possible to rewrite Eq.~\eqref{eq:Lindblad_equation_gen} as
\begin{align}
    \hat{\mathcal{L}} \left[ K \right] \rho(t)
&= 
    \frac{\Gamma}{2} \bm{\sigma} \cdot \bigg[ 2i \bm{k} \times \bm{k}^{*} + \left( \bm{k} \cdot \bm{n}(t) \right) \bm{k}^{*} \label{eq:2by2_Lindblad_Eq_in_n_k} \\
&+ 
    \left( \bm{k}^{*} \cdot \bm{n}(t) \right) \bm{k} - 2 \vert \bm{k} \vert^2 \bm{n}(t) \bigg], \nonumber
\end{align}
As the steady state $\rho_{\rm s}$ should satisfy $d\rho_{\rm s}/dt = \hat{\mathcal{L}} [K]\rho_{\rm s} = 0$, we can use Eq.~\eqref{eq:2by2_Lindblad_Eq_in_n_k} to express the corresponding Bloch vector $\bm{n}_{s}$ in terms of the components of the vector $\bm{k}$,
\begin{align}
    & \bm{n}_{s} = - \frac{2}{ \vert \bm{k} \vert^2 }\left(
     k_y k_z \sin (\phi_{yz}), \; k_z k_x \sin (\phi_{zx}), \; k_x k_y \sin (\phi_{xy}) 
    \right), \label{BV:2by2_general_steady_state}  
\end{align}
with $\phi_{jk} = \phi_{j} - \phi_{k}$. Next, we need to find how to ensure the existence of a chiral symmetry, which brings about a topological order in the steady state.

\subsection{Topological order of the steady-state density matrix} \label{subsec:2by2_Topological_order}

To be useful for quantum computation, the steady state should be pure, i.e., $\rho^{2}_{\rm s} \equiv \rho_{\rm s}$. According to Eq.~\eqref{DM:2by2_general_form_Bloch_vector}, this corresponds to $\bm{n}_s$ being a unit vector. A sufficient condition for the steady-state Bloch vector \eqref{BV:2by2_general_steady_state} to have unit length is
\begin{align}
    k_y^2 = -k_x^2 e^{\pm2 i (\phi_y - \phi_x )} -k_z^2 e^{ \pm 2 i (\phi_y  - \phi_z)}. \label{eq:2by2_ky_solution_unit_BV}
\end{align}
As the values of $k_{x}$, $k_{y}$ and $k_{z}$ must be real, only discrete values are allowed for the differences of the phases $\phi_{j}$. Since in an adiabatic transformation it should be possible to vary the parameters along a continuous path, we choose the phases of the vector $\bm{k}$ as follows,
\begin{align}
    \phi_{x} - \phi_{y} = \frac{ \pi }{2}, \qquad \phi_{z} - \phi_{y} = \frac{ \pi }{ 2 }. \label{eq:2by2_pure_state_condition_ky_phi}
\end{align}
These conditions lead to the following vector $\bm{k}$ and the corresponding steady-state Bloch vector $\bm{n}_{s}$,
\begin{align}
    \bm{k} &= \left( i k_x, \text{sgn}(k_y) \sqrt{k_x^2 + k_z^2}, i k_z \right) e^{i \phi_y }, \label{eq:2by2_vector_k_pure_state} \\
    \bm{n}_{s} &= \left( \frac{k_z}{k_y} , 0, - \frac{k_x}{k_y}  \right). \label{BV:2by2_pure_state}
\end{align}
One should note that there is still one free phase $\phi_{y}$ in the vector $\bm{k}$, but this does not affect the Bloch vector $\bm{n}_s$. 

\begin{figure}[t]
	\includegraphics[width=0.85 \columnwidth]{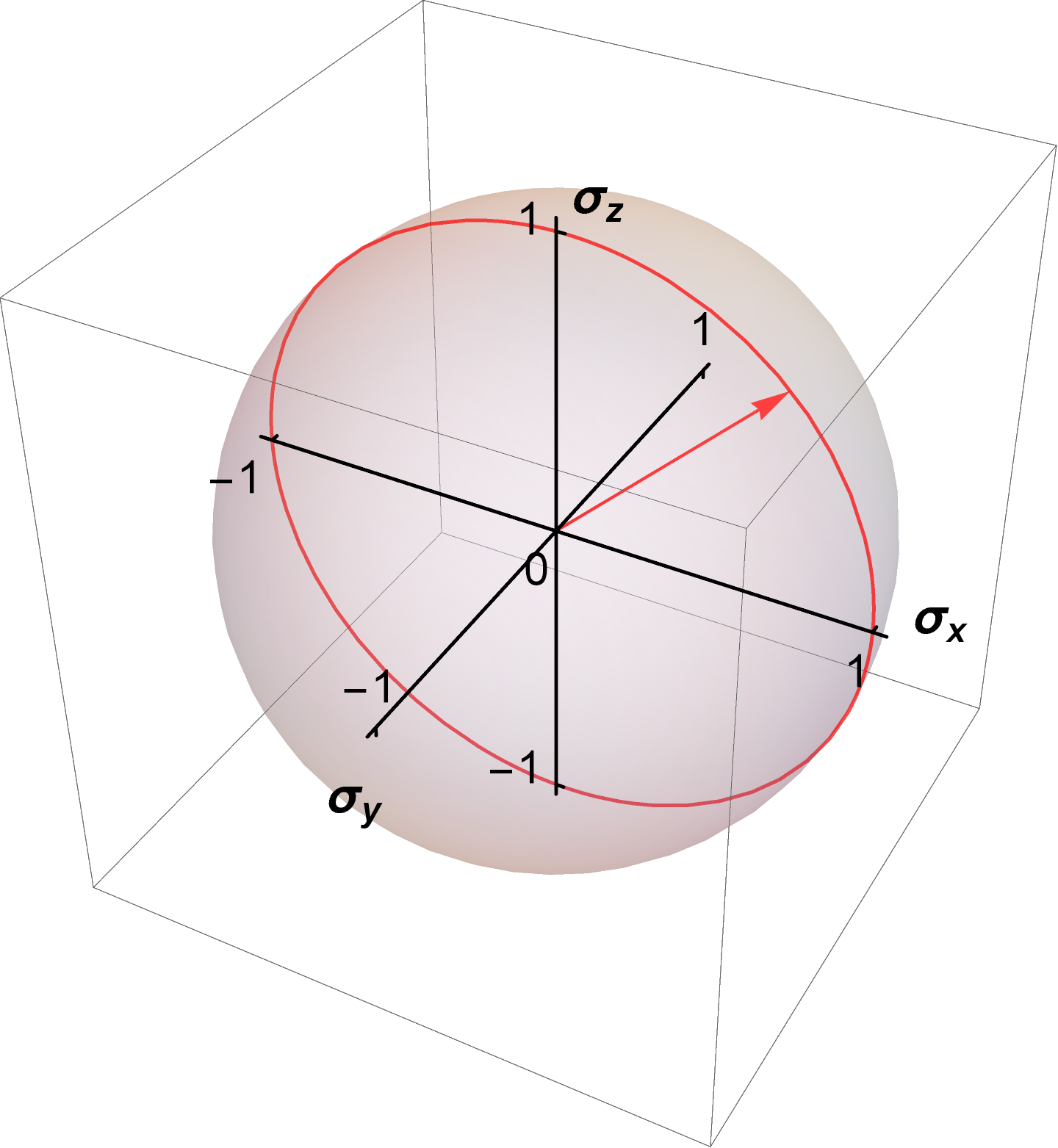}
	\caption{\label{fig:BlochSphere_BV_in_xzPlane} (color online) The Bloch vector \eqref{BV:2by2_pure_state} of a pure state is the red arrow in the Bloch sphere. While we rotate the qubits by changing parameters adiabatically, this Bloch vector also winds around the red solid unit circle in the $\sigma_{x} - \sigma_{z}$ plane.} 
\end{figure}

The steady state being pure is one of the conditions for rotations of the qubit state to give rise to a topological phase. The second condition is the existence of a chiral symmetry \cite{Zoller2011}. This latter exists if it is possible to find a unitary symmetry operator $\Sigma$ such that
\begin{align}
    \Sigma \left( \bm{n} \cdot \bm{\sigma} \right) \Sigma^{\dagger} = - \left( \bm{n} \cdot \bm{\sigma} \right) \; \Leftrightarrow \; \left\{ \Sigma, \; \bm{n} \cdot \bm{\sigma} \right\} = 0. \label{eq:chiral_symmtry_operator}
\end{align}
Writing $\Sigma = \bm{a} \cdot \bm{\sigma}$ with a unit vector $\bm{a}$, Eq.~\eqref{eq:chiral_symmtry_operator} is equivalent to the Bloch vector $\bm{n}$ and the vector $\bm{a}$ being orthogonal, i.e., $\bm{a} \cdot \bm{n} = 0$. Hence, the Bloch vector can only rotate on the Bloch sphere in a plane with normal vector $\bm{a}$. For the family of Bloch vectors in Eq.~\eqref{BV:2by2_pure_state}, the chiral symmetry operator is given by the pure-state Bloch vector $\bm{a} = (0, 1, 0)$ because $\bm{n}$ lies in the $y$ plane (see Fig.~\ref{fig:BlochSphere_BV_in_xzPlane}).

As a consequence, adiabatically rotating the Bloch vector \eqref{BV:2by2_pure_state} between time $t = 0$ and time $t = T$ along the circle shown in Fig.~\ref{fig:BlochSphere_BV_in_xzPlane} spans the solid angle
\begin{align}
    \Omega &= \int_{0}^{T} dt  \left[ \bm{n} (k_x, k_z) \times \partial_t \bm{n} (k_x, k_z) \right] \cdot \bm{a} \label{SA:2by2_general_xz_plane} \\
&= \int_{0}^{T} dt \; \frac{k_x \partial_t k_z - k_z \partial_t  k_x}{k_z^2 + k_x^2}. \label{SA:2by2_case_ky}
\end{align}
By writing $n_{x}$ and $n_{z}$ in polar coordinates as $n_{x}(t) = \sin \Theta(t)$ and $n_{z} (t) = \cos \Theta (t)$, we encode the adiabatic time evolution into the angle $\Theta(t)$. Then, the solid angle becomes (see also App.~\ref{app:solid_angle}).
\begin{align}
    \Omega &= \int_{0}^{T} dt \; \partial_t \Theta(t) = \Theta(T) - \Theta(0). \label{SA:general_in_Theta}
\end{align}
As $\Theta = \arctan \left( n_{x} / n_{z} \right) = -\arctan \left( k_{z} / k_{x} \right)$, we conclude that the solid angle only depends on the initial and final positions of the Bloch vector.

Hence, a cyclic adiabatic change of parameters leads to a winding number $\Omega = 2 \pi n$ with $n \in \mathbb{Z}$. It is topological in that it is quantized and independent of the exact time-dependent function chosen for the adiabatic change of parameters. The next step is to determine the winding number explicitly for the driven-dissipative Majorana system.

\subsection{The physical tunneling system for a single Majorana box} \label{subsec:tunneling_system_single_MB}

We now consider the tunneling couplings as shown in Fig.~\ref{fig:2by2_setup01}. By taking into account all possible transport processes from QD1 to QD2, we can find the following jump operator using Eq.~\eqref{eq:operator_A},
\begin{align}
    K_{12} 
&= \left( k_x e^{i \phi_x},k_y e^{i \phi_y},k_z e^{i \phi_z} \right) \cdot \bm{\sigma}, \label{eq:2by2_jumpoperator_arbitrary}
\end{align}
where
\begin{align}
k_x e^{i \phi_x} &= \lambda_{14} \lambda_{22} e^{-i \beta_{14} + i \beta_{22} - i \pi / 2 },  \label{eq:2by2_kx_single_MB_setup01} \\ 
k_y e^{i \phi_y} &= \lambda_{14} \lambda_{21} e^{-i \beta_{14} + i \beta_{21} - i \pi / 2 }, \label{eq:2by2_ky_single_MB_setup01} \\
k_z e^{i \phi_z} &= \lambda_{14} \lambda_{23} e^{-i \beta_{14} + i \beta_{23} - i \pi/2 }. \label{eq:2by2_kz_single_MB_setup01} 
\end{align}
The steady-state Bloch vector should be of the form~\eqref{BV:2by2_general_steady_state}. To stabilize such a pure state, the tunneling parameters should fulfill the conditions in Eq.~\eqref{eq:2by2_pure_state_condition_ky_phi}, which lead to
\begin{align}
& \beta_{22} = \beta_{23} = \beta_{21} + \frac{\pi}{2},  \label{eq:2by2_DD_protocol_beta} \\
& \lambda_{21}^2 = \lambda_{22}^2 + \lambda_{23}^2. \label{eq:2by2_DD_protocol_lambda}
\end{align}
In this case, we obtain the following results for the vector $\bm{k}$ and the pure-state Bloch vector $\bm{n}_s$,
\begin{align}
    \bm{k}
&= 
    \left( i \lambda_{22}, \text{sgn}(\lambda_{21}) \sqrt{ \lambda_{22}^2 + \lambda_{23}^2},i \lambda_{23} \right) \lambda_{14} e^{ i \Phi_{y} } , \label{eq:2by2_vector_k12_setup01} \\
    \bm{n}_s 
&= 
    \frac{ \text{sgn}(\lambda_{21}) }{\sqrt{\lambda_{23}^2 + \lambda_{22}^2}} \left( \lambda_{23},0,- \lambda_{22} \right), \label{BV:2by2_pure_state_single_MB_setup01} 
\end{align}
where $\Phi_{y} = \beta_{21} - \beta_{14} - \pi/2$.

If we consider again a time-periodic adiabatic change of the parameters $\lambda_{22}(t)$ and $\lambda_{23}(t)$, we can again introduce polar coordinates such that their time-dependence is indicated by the azimuthal angle $\varphi$,
\begin{align}
    \lambda_{22}(t) = - r \cos \varphi(t),\; \lambda_{23} = r \sin \varphi(t) . \label{eq:2by2_polar_coordinate_lambda}
\end{align}
The steady-state Bloch vector \eqref{BV:2by2_pure_state_single_MB_setup01} thus evolves in time as $\bm{n}_s(t) = \text{sgn}(\lambda_{21}) ( \sin \varphi(t), 0, \cos \varphi(t) )$. Inserting this Bloch vector into Eq.~\eqref{SA:general_in_Theta} gives the quantized solid angle of $ \Omega = 2 \pi$ for a single loop in parameter space.

\subsection{Summary}

We started this section by considering a general expression for the steady-state Bloch vector~\eqref{BV:2by2_general_steady_state} of a Majorana box coupled to two QDs, in the case when one transition rate is dominant. To drive the system to a pure steady state, we can choose a parameter set satisfying the condition~\eqref{eq:2by2_pure_state_condition_ky_phi}. This choice leads to the existence of a chiral symmetry~\eqref{eq:chiral_symmtry_operator} in the steady state, so rotating the Bloch vector~\eqref{BV:2by2_pure_state} on the Bloch sphere can yield a quantized winding number. It is given by the solid angle~\eqref{SA:2by2_case_ky} spanned during the time evolution of the Bloch vector. The realization of the topological order in this reduced parameter space arises from the fact that the winding number is quantized and does not depend on the trajectory of the parameters over time. 

We showed how a topological phase can arise from adiabatic changes in the parameters of a driven-dissipative Majorana box qubit. However, the requirement that the stabilized dark state should be a pure state constrains the parameter ranges for the allowed tunnel couplings. In addition, the parameter space needs to allow for the definition of a chiral symmetry operator. The reduced parameter space leaves sufficient degrees of freedom to allow an adiabatic rotation of the qubit state vector which gives rise to a topological winding number.

For braiding, however, we will need multiple dark states in the same parity subspace, whereas the dark state subspace of a single Majorana box has at most two dark states of a given parity. Therefore, in the subsequent section, we will consider the stabilization and rotation of a qubit state in a system consisting of more than one coupled Majorana box and develop a braiding protocol for that system.

\section{Braiding in the decoherence-free subspace by adiabatic parameter changes} \label{sec:braiding_in_DFS}

A braiding process in this driven-dissipative system can be realized by a change of steady state from one dark state to another one in the same parity subspace, while the system parameters are changed adiabatically along a closed path. The following requirements should be fulfilled to allow such braiding:
\begin{enumerate}
    \item The dark state subspace should contain multiple dark states with the same parity.
    \item The dark states should be parameter-dependent so that they can be reached by adiabatically changing parameters.
    \item Dark states do not disappear and do not cross when parameters are slowly tuned during braiding.
\end{enumerate}
In the Lindblad equation $\dot{\rho} = \mathcal{L}[K] \rho$, a dark state can be found as the eigenstate of the jump operator with eigenvalue zero \cite{Bardyn2013}. If a state $\ket{\psi}$ satisfies $K \vert \psi \rangle = 0$, this implies that $\mathcal{L}[K] \vert \psi \rangle \langle \psi \vert = 0$, so that $\rho = \vert \psi \rangle \langle \psi \vert$ is the pure steady state of the system. In this section, we begin by considering an ideal jump operator whose dark states can fulfill the above three requirements. Next, we propose a geometry of tunneling couplings which realizes this jump operator and uses it to derive a braiding protocol. 

\subsection{The ideal jump operator} \label{subsec:three_MBs_ideal_jump_operator}

In a braiding transformation, an adiabatic change of system parameters along a closed path in parameter space brings the system from an initial state $\rho_i$ to a final state $\rho_f = U \rho_i$, where $U$ is the braiding matrix. If we prepare an initial state with a given parity and use a jump operator which does not couple different parity sectors, the two parity blocks in the density matrix will remain separated during the braiding protocol, i.e., $\mathcal{L}[K] \rho = \mathcal{L}[K_{\rm even}] \rho_{\rm even} \otimes \mathcal{L}[K_{\rm odd}] \rho_{\rm odd}$. Since the two parity sectors are uncoupled, we will consider only the odd parity sector for the rest of this section.

As described in the first requirement, we need at least two Majorana qubits to obtain two dark states with the same parity. In this case, the jump operator $K_{\rm odd}$ of two coupled Majorana boxes is a $2 \times 2$ matrix in the odd-parity subspace spanned by the states $\{ \ket{01}, \ket{10} \}$. For a general $2 \times 2$ jump operator, we can use the expression in Eq.~\eqref{eq:2by2_jumpoperator_arbitrary}. Its eigenvalues and the corresponding eigenstates are 
\begin{align}
    K_{\rm odd} \ket{k_{\pm}} = \pm |\bm{k}| \ket{ k_{\pm}}, \label{eq:4by4_eigenvalues_jump_operator_Kodd}
\end{align}
with $\ket{k_{\pm}} = (k_{z} e^{ i \phi_{z} } \pm | \bm{k}| , \; k_{x} e^{ i \phi_{x} } + k_{y} e^{ i \phi_{y} } )$. A nonzero $K_{\rm odd}$ has a single dark state $(k_{z} e^{ i \phi_{z} }, k_{x} e^{ i \phi_{x} } + k_{y} e^{ i \phi_{y} } )$. Therefore, this setup does not provide sufficient ingredients for braiding, namely \emph{multiple} parameter-dependent dark states in the same-parity subspace. Therefore, we need to extend the Hilbert space by adding one more Majorana box.

In the case of coupled three Majorana boxes, the odd-parity subspace is four-dimensional, so we assume that $K_{\rm odd}$ is a $4 \times 4$ matrix in the odd-parity state manifold spanned by $\{ \ket{ 001},  \ket{ 010 }, \ket{ 100 },  \ket{ 111 } \}$. A possible choice for the jump operator, which will turn out to be realizable in the Majorana box system, is
\begin{align}
    K_{\rm odd} = \begin{pmatrix}
	0 & f_{1} & - f   f_{1} & 0 \\
	g_{1} & 0 & 0 & g   g_{1} \\
	g_{2} & 0 & 0 & g   g_{2} \\
	0 & f_{2} & - f   f_{2} & 0
\end{pmatrix}, \label{eq:8by8_jump_operator_f_g_form}
\end{align}
where $f$ and $g$ are real functions of the tunnel couplings, $f(\lambda_{jv}, \beta_{jv})$ and $g(\lambda_{jv},  \beta_{jv})$, such that the adiabatic time evolution is encoded in $f$ and $g$. This jump operator gives rise to the two dark states, $K_{\rm odd} \ket{\psi_{1,2}} = 0$, with
\begin{align}
\ket{\psi_{1}} &= N_1 \left( -g \ket{001} + \ket{111} \right), \label{eq:8by8_darkstate_psi1_f_g_form} \\
\ket{\psi_{2}} &= N_2 \left( f \ket{010} + \ket{100} \right), \label{eq:8by8_darkstate_psi2_f_g_form}
\end{align}
where $N_1$ and $N_2$ are normalization constants. These two dark states are independent of $f_{1,2}$ and $g_{1,2}$, yet they are distinct when $g_{1}$ and $g_{2}$ are non-zero and $f_{1}$ and $f_{2}$ are non-zero at the same time. Moreover, they depend on the tunneling parameters through $f$ and $g$, which are required for braiding in the dark state subspace. They do not coalesce for any values of $f$ and $g$.

The other two eigenvalues of $K_{\rm odd}$ and their corresponding eigenvectors are, respectively,
\begin{align}
    K_{\rm odd} \ket{ \psi_{\pm} } 
&= 
    \pm \sqrt{ \left( f_{1}+f_{2} g \right) \left( g_{1} - f g_{2} \right) } \ket{ \psi_{\pm} } \label{eq:8by8_two_other_eigenvalues_of_jump_operator_f_g_form}  \\
    \ket{ \psi_{\pm} } &= N_3 \begin{pmatrix}
       f_{1} \sqrt{g_{1}-f g_{2}} \\
       \pm g_{1} \sqrt{f_{1}+f_{2} g} \\
        \pm g_{2} \sqrt{f_{1}+f_{2} g} \\
        f_{2} \sqrt{g_{1}-f g_{2}}
    \end{pmatrix}
\end{align}
with a normalization $N_{3}$. Since the eigenvalues~\eqref{eq:8by8_two_other_eigenvalues_of_jump_operator_f_g_form} also depend on the parameters $f$ and $g$, it is possible that they become zero when $f_{1}$=$-f_{2} g$ or $g_{1}$=$f g_{2}$ during the adiabatic change in parameters. In that case, the eigenvectors $\ket {\psi_{\pm} } $ will coalesce with one of the dark states, but this does not affect the braiding protocol because the two dark states still remain distinct.

Hence, the jump operator~\eqref{eq:8by8_jump_operator_f_g_form} constitutes a viable option for braiding, and the next step is to realize this operator in the Lindblad equation~\eqref{eq:Lindblad_equation_2QDs}, such that the steady state will be given by the two dark states $\ket{ \psi_{1} }$ and $\ket{\psi_{2}}$. 

\subsection{The pure steady state and its topological order}
\label{subsec:stabilization_in_three_MBs}

We start again from the Lindblad equation~\eqref{eq:Lindblad_equation_2QDs} and consider again the weak driving regime as in Sec.~\ref{subsec:2by2_DM_SS_general_form}. In this case $\Gamma_{12} \gg \Gamma_{21}$, so we consider $\Gamma_{12}$ as the only decay rate. The Lindblad term $\mathcal{L}[K_{\rm odd}]$ with the jump operator~\eqref{eq:8by8_jump_operator_f_g_form} then drives the Majorana qubits to a steady state $\rho_s$ obeying $\mathcal{L}[K_{\rm odd}] \rho_{\rm s} = 0$. 

As we show in App.~\ref{app:three_boxes_Lindbladian_stabilization}, we can use the following ansatz for the steady state density matrix,
\begin{align}
    \rho_{\rm s} &= \sum_{n = 0} ^{3} a_{nn} \ket{n}\bra{n}  + \sum_{n> m} \left[ \left( a_{nm} + i b_{nm} \right) \ket{n} \bra{m} + \text{h.c.}\right], \label{eq:8by8_arbitrary_DM}
\end{align}
where $a_{nm}, b_{nm} \in \mathbb{R}$ and the states $\ket{0}$, $\ket{1}$, $\ket{2}$ and $\ket{3}$ are shortcuts for the odd-parity states $\ket{001}$, $\ket{010}$, $\ket{100}$ and $\ket{111}$, respectively. Due to the properties of a density matrix, we assume $a_{nn} \geq 0$ and $\sum_{n} a_{nn} \equiv 1$ for all $n$. Using the basis vectors $\vert \psi_{1} \rangle$ and $\vert \psi_{2} \rangle$ from Eqs.~\eqref{eq:8by8_darkstate_psi1_f_g_form}-\eqref{eq:8by8_darkstate_psi2_f_g_form} as well as two orthonormal vectors $\vert \psi_{3} \rangle$ and $\vert \psi_{4} \rangle$,
\begin{align}
    \vert \psi_{3} \rangle &= \frac{1}{ \sqrt{ 1+ g^2} } \left( \vert 001 \rangle + g \vert 111 \rangle \right)  \label{eq:psi_3} \\
    \vert \psi_{4} \rangle &= \frac{1}{ \sqrt{ 1+ f^2} } \left( - \vert 010 \rangle + f \vert 100 \rangle \right). \label{eq:psi_4}
\end{align}
one finds that the pure steady state $\rho_{\rm s}$ can be constructed from the dark states $\vert \psi_{1} \rangle$ and $\vert \psi_{2} \rangle$ and can be written as,
\begin{align}
    \rho_{\rm s} = \left( \sqrt{1-F^{2}} e^{i\alpha_{1}} \vert \psi_{1} \rangle +   F e^{i\alpha_{2}} \vert \psi_{2} \rangle \right) 
    \otimes \text{h.c.} , \label{DM:8by8_pure_state_F_alpha12}
\end{align}
where the following two conditions have to be satisfied (see Eq.~\eqref{appeq:8by8_steady_state_a03_a12} in App.~\ref{app:three_boxes_Lindbladian_stabilization}),
\begin{align}
    F^2 = \frac{1+f^2}{ f } a_{12} \ \text{ and } \ F^2  =  1 +  \frac{1+g^2}{ g } a_{03}, \label{eq:8by8_steady_state_a03_a12}
\end{align}
and where $a_{12}$ and $a_{03}$ are the real parts of the elements at $\vert 010 \rangle \langle 101 \vert$ and $\vert 001 \rangle \langle 110 \vert$ in the general form of the steady state \eqref{eq:8by8_arbitrary_DM}, respectively. Moreover, the phases $\alpha_{1}$ and $\alpha_{2}$ are associated to the argument of the element at $\vert 100 \rangle \langle 110 \vert$ by $\arctan(b_{23}/a_{23}) = \alpha_{1} - \alpha_{2}$.

The same result can also be proven by writing the Liouvillian superoperator $\mathcal{L}$ in the form of a matrix \cite{Gau2020}. After some algebra, one can find four dark states corresponding to the vectorized forms of $\vert \psi_{l} \rangle \langle \psi_{m} \vert$, with $l$, $m$ $\in$ $\{1, 2\}$. The other eigenstates have negative eigenvalues, such that they decay exponentially over time.

To ensure the chiral symmetry required for the existence of a topological phase, we focus on linear combinations of two dark states with equal phases, $\alpha_{1} = \alpha_{2}$, and obtain the following density matrix from Eq.~\eqref{DM:8by8_pure_state_F_alpha12} (see also App.~\ref{app:three_boxes_Lindbladian_stabilization}),
\begin{align}
    \rho_{\rm s}
&=  \left( 1 -F^{2} \right) \vert \psi_{1} \rangle \langle \psi_{1} \vert + F^{2} \vert \psi_{2} \rangle \langle \psi_{2} \vert \nonumber \\
    &+ F \sqrt{1 - F^{2}} \left( \vert \psi_{1} \rangle \langle \psi_{2} \vert + \vert \psi_{2} \rangle \langle \psi_{1} \vert\right). \label{DM:8by8_steady_state_pure}
\end{align}
Considering this density matrix as a vector on the Bloch sphere spanned by $\left\{ \vert \psi_{1} \rangle, \vert \psi_{2} \rangle \right\}$, we can again write $\rho_{\rm s} = \left( \sigma_{0} + \bm{n} \cdot \bm{\sigma} \right)/2$ with
\begin{align}
    \bm{n} 
&= 
    \left( 2  F \sqrt{ 1 - F^2 }, \; 0, \; 1 - 2 F^2 \right). \label{BV:8by8_bloch_vector_in_terms_of_F}
\end{align}
The chiral symmetry is again implemented by $\Sigma = \sigma_{y}$, and upon an adiabatic change of parameters, the Bloch vector~\eqref{BV:8by8_bloch_vector_in_terms_of_F} rotates in the $\sigma_{x}$-$\sigma_{z}$ plane. This is analogous to Fig.~\ref{fig:BlochSphere_BV_in_xzPlane}, but the Pauli matrices here are in the space of dark states $\{ \ket{ \psi_{1} }, \ket{ \psi_{2}}\}$.

As in Sec.~\ref{subsec:2by2_Topological_order} the solid angle $\Omega$ is given by the winding number of the Bloch vector~\eqref{BV:8by8_bloch_vector_in_terms_of_F} between time $t_{0}$ to $t_{f}$ (see App.~\ref{app:solid_angle}), 
\begin{align}
    \Omega  &= \Theta[f(t_{f}), g(t_{f})] - \Theta[f(t_{0}), g(t_{0})], \label{SA:8by8_in_form_of_Theta} \\
    \Theta &= \arctan \left( \frac{ n_{x} }{ n_{z} } \right) = \arctan \left( \frac{ 2 F \sqrt{ 1-F^{2} } }{ 1- 2 F^{2}} \right) \nonumber.
\end{align}
The tunnel couplings are encoded in the functions $f$ and $g$, whose periodic change can be expressed using polar coordinates with an angle $\theta(t_{f})= \theta(t_{0}) + 2\pi$.
\begin{align}
    f(t) = r \cos[\theta(t)] \; \text{ and } \; g(t) = r \sin[\theta(t)], \label{eq:8by8_f_g_polar_coordinate}
\end{align}

For a nontrivial braiding process, the Bloch vector~\eqref{BV:8by8_bloch_vector_in_terms_of_F} should wind about a semicircle while the parameters are driven along a closed loop back to their initial values. We therefore choose the function $F(f, g) \in [-1,1]$ as follows,
\begin{align}
F[\theta(t)] = \sin \left[ \frac{\theta(t)}{4} \right] = \sin \left[ \frac{1}{4} \arcsin \left( \frac{g}{\sqrt{ f^2 + g^2 }} \right) \right]. \label{eq:8by8_case_pi_F}
\end{align}
In this case, the Bloch vector \eqref{BV:8by8_bloch_vector_in_terms_of_F} becomes $\bm{n} = ( \sin(\theta/2), 0, \cos(\theta/2) )$ and the solid angle~\eqref{SA:8by8_in_form_of_Theta} reaches $\Omega = \pi$ after driving $f$ and $g$ for one period,
\begin{align}
    \theta(t_{0})&=0 & \bm{n}(t=0) &= \left( 0, 0, 1 \right), \notag \\
    \theta(t_{f})&=2 \pi & \bm{n}(t=T) &= \left( 0, 0, -1\right). \label{eq:8by8_braiding_process_in_Bloch_vector}
\end{align}
The corresponding braiding operation can be expressed as a braiding operator $B = \sigma_y$ acting on the Hilbert space spanned by the two dark states~\eqref{eq:8by8_darkstate_psi1_f_g_form}-\eqref{eq:8by8_darkstate_psi2_f_g_form},
\begin{align}
& B \left[ \bm{n}(t= 0) \cdot \bm{\sigma} \right] B^{\dagger} = \bm{n}(t= T) \cdot \bm{\sigma} \notag \\ 
& \Rightarrow  B \sigma_{z} B^{\dagger} = - \sigma_{z}. \label{eq:8by8_braiding_operator_sigmay}
\end{align}
With the ideal jump operator $K_{\rm odd}$ in Eq.~\eqref{eq:8by8_jump_operator_f_g_form}, the Majorana qubits are stabilized in the dark state subspace where we can carry out braiding in the odd-parity subspace by tuning the tunneling parameters. To achieve this, it is necessary to prepare an odd-parity initial state of the system and the phase coherence between two dark states should ensure the chiral symmetry in the steady state \eqref{DM:8by8_steady_state_pure} (see App.~\ref{app:three_boxes_Lindbladian_stabilization}). The next step is to find a tunneling system whose jump operator provides the structure of $K_{\rm odd}$ in Eq.~\eqref{eq:8by8_jump_operator_f_g_form}, and then demonstrate a braiding protocol in this open system.

\subsection{The tunneling system and the braiding protocol} \label{subsec:three_MBs_tunneling_system_and_braiding_protocol}

In this section, we consider the same open quantum system as depicted in Fig.~\ref{fig:2by2_setup01}, but with two additional Majorana boxes. Analogously to Eq.~\eqref{Hamil:the_entire_system}, the total Hamiltonian is 
\begin{align}
    H(t) = H_{\rm box} + H_{\rm QD} + H_{\rm env} + H_{\rm drive}(t) + H_{\rm tun, e} + H_{\rm tun, \gamma}, \label{Hamil:8by8_general_form}
\end{align}
where $H_{\rm box}$, $H_{\rm QD}$, $H_{\rm env}$, and $H_{\rm drive}(t)$ represent the Hamiltonian of the Majorana box, quantum dots, environment, and the time-dependent driving field, respectively. For three Majorana boxes, $H_{\rm box} = E_{C} \sum_{n=1}^{3} (N_{n} - N_{g})^2$. The tunneling Hamiltonian consists of two contributions. Electron tunneling between the quantum dots and Majorana box $n$ is represented by
\begin{align}
H_{\rm tun,e} &= t_{0} \sum_{n=1}^3 \sum_{ j=1}^2 \sum_{\mu=1}^4 \lambda_{j, n \mu} e^{ -i \beta_{j, n \mu} } e^{-i\hat{\phi}_{n} + i\delta_{j}} d_{j}^{\dagger} \gamma_{n \mu} + \text{h.c.}, \label{Hamil:8by8_tunneling_Hamiltonian} 
\end{align}
while direct tunneling between the two Majorana boxes (the box $n$ and the box $n'$) is described by
\begin{align}
H_{\rm tun,\gamma} &= E_{C} \sum_{n,n' = 1}^3 \sum_{\mu, \nu =1}^4 i \tilde{t}_{n n'} \gamma_{n \mu} \gamma_{n' \nu}, \label{Hamil:8by8_tunnels_between_MBs} 
\end{align}
with the dimensionless parameter $\tilde{t}_{n n'} = t_{n n'} / E_{C}$, and $t_{nn'}$ is the tunnel coupling energy between the boxes $n$ and $n'$. Here, we assume that $t_{nn} = 0$ and $t_{n n'} > 0$ for all amplitudes.

In analogy to Sec.~\ref{subsec:dissipation_in_Hamiltonian}, we apply the RWA and the SW transformation (see App.~\ref{app:three_MBs_jump_operator_setup01}). 
In the SW transformation, the small perturbations are the couplings between QDs and MBSs $t_{0}$ as well as those between the Majorana boxes $t_{nn'}$. For three coupled Majorana boxes, the SW transformation should be expanded up to the fourth order to obtain the cotunneling Hamiltonian in the low-energy regime. For large $E_C$, the result can be approximated as
\begin{align}
    H_{\rm cot} \approx \sum_{r=2}^{4} H_{\rm cot}^{(r)}= \sum_{r=2}^{4} \left( \frac{1}{(r-1)!} - \frac{1}{r!} \right)\left( \frac{2}{ E_{C} } \right)^{r-1} \left( H_{\rm tun} \right)^{r}, \label{Hamil:8by8_cotunneling_Hamiltonian}
\end{align}
where $H_{tun} = H_{tun, e} + H_{tun, \gamma}$. 

\begin{figure}[t]
    \centering
\begin{tikzpicture}[every text node part/.style={align=center}]

    \begin{pgfonlayer}{foreground}
    \filldraw [blue] (2.5, 1.0) circle (2pt) node[blue, anchor=west] at (2.55, 1.2) {\large QD $1$};

    \draw[dotted, blue, very thick] (2.5, 1.0) -- (-0.1, -0.15) 
    ;
    \draw[dotted, blue, very thick] (2.5, 1.0) -- (-0.45, -0.78) 
    ;

    \draw[dotted, blue, very thick] (2.5, 1.0) -- (4.1 +1, -0.15) 
    ;
    \draw[dotted, blue, very thick] (2.5, 1.0) -- (4.45 +1, -0.78) 
    ;

    \draw[black, thick] (-0.1, -0.15) node[cross=3pt, rotate=90]{} node[anchor=west] at (-0.6, 0.1 + 0.3) {\large $\bm{\gamma_{11}}$};
    \draw[black, thick] (1.4, -0.95) node[cross=3pt, rotate=90]{} node[anchor=west] at (1.0 +0.55, -0.7 + 0.15) {\large $\bm{\gamma_{13}}$};
    \draw[black, thick] (-0.45, -0.78) node[cross=3pt, rotate=90]{} node[anchor=west] at (-0.95-0.15, -0.53-0.8){\large $\bm{\gamma_{12}}$};
    \draw[black, thick] (1.05, -1.58) node[cross=3pt, rotate=90]{} node[anchor=west] at (0.64 -0.15, -1.33-0.8) {\large $\bm{\gamma_{14}}$};

    \draw[black, thick] (4.1 +1, -0.15) node[cross=3pt, rotate=90]{} node[anchor=east] at (4.6 +1, 0.1+0.3) {\large $\bm{\gamma_{22}}$};
    \draw[black, thick] (2.6 +1, -0.95) node[cross=3pt, rotate=90]{} node[anchor=east] at (3.0 +1 -0.55, -0.7+0.15){\large $\bm{\gamma_{24}}$};
    \draw[black, thick] (4.45 +1, -0.78) node[cross=3pt, rotate=90]{} node[anchor=east] at (4.95 +1 + 0.15, -0.53 -0.8){\large $\bm{\gamma_{21}}$};
    \draw[black, thick] (2.95 +1, -1.58) node[cross=3pt, rotate=90]{} node[anchor=east] at (3.36 +1 +0.15, -1.33 -0.8){\large $\bm{\gamma_{23}}$};

    \filldraw [red] (2.5, -5.0) circle (2pt) node[red, anchor=west] at (2.55, -5.2) {\large QD $2$};
    \filldraw [color=black, fill=white, thick] (-1.1, -3.5) circle (0.35) node[anchor=west] at (0.575-2.1, -3.5) {\large AC};
    \draw[dotted, red, very thick] (2.5, -5.0) -- (2.5-0.212-0.142, -2.5 -1.70) 
    ;
    \draw[dotted, red, very thick] (2.5, -5.0) -- (2.5+0.212 +0.142, -2.5 -1.70) 
    ;
    \end{pgfonlayer}

    \draw[violet, thick] (1.4, -0.95) -- (2.6 +1, -0.95) node[violet] at (2.5, -1.2) {\large $t_{12}$};
    \draw[Magenta, thick] (2.95 +1, -1.58) -- (2.5 + 0.212 + 0.141, -2.5) node[Magenta] at (3.4015 - 0.2, -1.8) {\large $t_{23}$};
    \draw[RedOrange, thick] (2.05 -1, -1.58) -- (2.5 - 0.212 - 0.141, -2.5) node[RedOrange] at (5 -3.4015 + 0.2, -1.8) {\large $t_{13}$};

    \draw[black, thick] (2.5, 1.0) -- (-1.1, 1.0) -- (-1.1, -5.0) -- (2.5, -5.0);
    
    \begin{scope}[rotate around= {15 : (0, 0)}]
    \begin{pgfonlayer} {background}
    \filldraw[blue!17, rounded corners=10pt] (0, 0.4)  -- (1.6, -1.2) -- (0.5, -2.3) -- (-1.1, -0.7) -- cycle ;
    \begin{pgfonlayer} {background}
        \filldraw[YellowOrange, thick] (0.3, -0.7) -- (0.5, -0.9) -- (0.2, -1.2) -- (0, -1);
    \end{pgfonlayer}
    \end{pgfonlayer}
    \filldraw[cyan, thick] (0, 0) -- (1.2, -1.2) -- (1.0, -1.4) -- (-0.2, -0.2);
    \filldraw[cyan, thick] (-0.5, -0.5) -- (0.7, -1.7) -- (0.5, -1.9) -- (-0.7, -0.7);
    \end{scope}

    \begin{scope}[rotate around= {-15 : (4 +1, 0) }]
    \begin{pgfonlayer} {background}
        \filldraw[blue!17, rounded corners=10pt] (4 +1, 0.4) -- (2.4 +1, -1.2) -- (3.5 +1, -2.3) -- (5.1 +1, -0.7)  --cycle ;
    \begin{pgfonlayer}{background}
        \filldraw[YellowOrange, thick] (4.7, -0.7) -- (4.5, -0.9) -- (4.8, -1.2) -- (5, -1);
    \end{pgfonlayer}
    \end{pgfonlayer}
    \filldraw[cyan, thick] (4 +1, 0) -- (2.8 +1, -1.2) -- (3.0 +1, -1.4) -- (4.2 +1, -0.2);
    \filldraw[cyan, thick] (4.5 +1, -0.5) -- (3.3 +1, -1.7) -- (3.5 +1, -1.9) -- (4.7 +1, -0.7);
    \end{scope}

    \begin{pgfonlayer} {background}
        \filldraw[blue!17, rounded corners=10pt] (2.5-0.212-0.283 - 0.29, -2.5 + 0.29) rectangle (2.5+0.212+0.283 + 0.29, -2.5 -1.70 - 0.29) ;
        \begin{pgfonlayer}{background}
            \filldraw[YellowOrange, thick] (2.288, -2.5 - 0.7) rectangle (2.712, -4.2 + 0.7);
        \end{pgfonlayer}
    \end{pgfonlayer}
    \filldraw[cyan, thick] (2.5-0.212, -2.5)  rectangle (2.5-0.212-0.283, -2.5 -1.70);
    \filldraw[cyan, thick] (2.5+0.212, -2.5)  rectangle (2.5+0.212+0.283, -2.5 -1.70);

    \draw[black, thick] (2.5-0.212-0.141, -2.5) node[cross=3pt, rotate=90]{} node[anchor=east] at (2.5-0.212-0.141 -0.35, -2.5-0.2) {\large $\bm{\gamma_{32}}$};
    \draw[black, thick] (2.5-0.212-0.141, -2.5-1.70) node[cross=3pt, rotate=90]{} node[anchor=east] at (2.5-0.212-0.141 -0.35, -2.5-1.70 + 0.2){\large $\bm{\gamma_{34}}$};
    \draw[black, thick] (2.5 + 0.212 + 0.141, -2.5) node[cross=3pt, rotate=90]{} node[anchor=west] at (2.5+0.212+0.141 +0.35, -2.5-0.2) {\large $\bm{\gamma_{31}}$};
    \draw[black, thick] (2.5+0.212+0.141, -2.5-1.70) node[cross=3pt, rotate=90]{} node[anchor=west] at (2.5+0.212+0.141 +0.35, -2.5-1.70 + 0.2){\large $\bm{\gamma_{33}}$};

    \draw[black!70] node[anchor=east] at (0.5, -2.3 +0.3) { Majorana \\ box $1$};
    \draw[black!70] node[anchor=west] at (4.5, -2.3 +0.3) { Majorana \\ box $2$};
    \draw[black!70] node[anchor=east] at (1.8, -2.79 -0.55) { Majorana \\ box $3$};
      
\end{tikzpicture}

\caption{The proposed braiding setup consists of three Majorana boxes mutually linked by tunneling with amplitudes $t_{12}$, $t_{13}$ and $t_{23}$. The Majorana operator $\gamma_{n \mu}$ denotes the MBS $\mu \in \{1,\ldots,4\}$ in the box $n \in \{1,2,3\}$. Two QDs (QD1 and QD2) are coupled via tunneling to the MBSs. The quantum dots, the driving field, and the environment remain identical to the case of a single Majorana box. The tunneling couplings are designed for attaining the jump operator~\eqref{eq:8by8_jump_operator_f_g_form} whose odd-parity dark states depend on the parameters $f$ and $g$ as shown in Eqs.~\eqref{eq:8by8_darkstate_psi1_f_g_form}-\eqref{eq:8by8_darkstate_psi2_f_g_form}.}
\label{fig:8by8_setup01}
\end{figure}
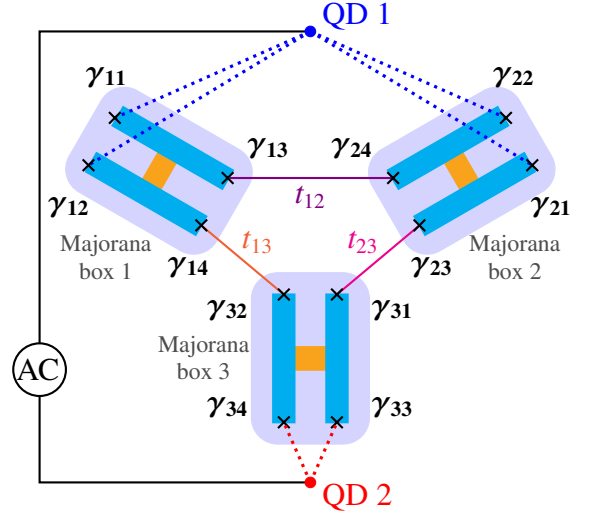

In this cotunneling Hamiltonian, we only retain terms describing complete tunneling trajectories between two dots because the remaining terms would only result in subleading corrections to higher-order tunneling processes. Hence, we find
\begin{align}
H_{\rm cot}^{(2)} & \approx \frac{ 1 }{E_{C}} H_{\rm tun, e} H_{\rm tun, e}, \label{Hamil:QD_QD_cotunneling_single_box} \\
H_{\rm cot}^{(3)} & \approx \frac{ 4 }{ 3 E_{C}^2 } H_{\rm tun, e} H_{\rm tun, \gamma} H_{\rm tun, e}, \label{Hamil:QD_QD_cotunneling_two_boxes} \\
H_{\rm cot}^{(4)} & \approx \frac{ 1 }{ E_{C}^3 } H_{\rm tun, e} H_{\rm tun, \gamma} H_{\rm tun, \gamma} H_{\rm tun, e} . \label{Hamil:QD_QD_cotunneling_three_boxes}
\end{align}
Thus, one can again obtain the effective Hamiltonian as follows.
\begin{align}
    H_{\rm eff} = H_{\rm QD} + H_{\rm env} + H_{\rm drive}^{RWA} + \sum_{r=2}^{4} H_{\rm cot}^{(r)}. \label{Hamil:effective_total_Hamiltonian_in_the_case_of_three_boxes}
\end{align}
This Hamiltonian serves as the starting point for the derivation of the Lindblad equation. Using the Born-Markov approximation, one can trace over the environment degrees of freedom and the quantum dot subspace if the stabilization time is long enough. As shown in App.~\ref{app:three_MBs_jump_operator_setup01}, one finds that the structure of the jump operator is associated with the Kronecker product of Majorana bilinears in each box,
\begin{align}
    & \mathcal{S}_{abc} = \mathcal{T} \cdot \left( \chi_{a}^{1} \otimes \chi_{b}^{2} \otimes \chi_{c}^{3} \right) \cdot \mathcal{T}^{-1}, \label{eq:8by8_tensorproducts_three_PauliMatrices}
\end{align}
for $a, b, c \in \{ 0, x, y, z\}$. Here, $\mathcal{T}$ represents a unitary matrix rearranging the blocks in matrices according to the parity and $\chi_a$ denotes sets of Pauli matrices. To obtain the same structure as $K_{\rm odd}$ in \eqref{eq:8by8_jump_operator_f_g_form}, App.~\ref{app:three_MBs_jump_operator_setup01} demonstrates the inclusion of $16$ necessary matrices $\mathcal{S}_{abc}$ within the tunneling system depicted in Fig.~\ref{fig:8by8_setup01}.

In App.~\ref{app:three_MBs_jump_operator_setup01}, we divide the jump operator of the setup in Fig.~\ref{fig:8by8_setup01} into four contributions $\tilde{K}_{\rm \rm odd}^{jk}$ ($j,k \in \{1,2\}$), with each superscript corresponding to a specific Majorana bound state as the starting point,
\begin{align}
    \tilde{K}_{\rm odd}^{11} &= \left[ \lambda_{2,  33} e^{i \beta_{2,  33} } \left( i \tilde{t}_{12} \tilde{t}_{23}  \mathcal{S}_{xzx}^{\rm odd} + i \tilde{t}_{13} \mathcal{S}_{y0y}^{\rm odd} \right) \right. \label{eq:8by8_K_odd_11} \\
     & \left. + \lambda_{2,  34} e^{i \beta_{2,  34}} \left( i \tilde{t}_{12} \tilde{t}_{23} \mathcal{S}_{xzy}^{\rm odd} - i \tilde{t}_{13} \mathcal{S}_{y0x}^{\rm odd} \right) \right] \lambda_{1,  11} e^{-i \beta_{1,  11} } , \nonumber \\
    \tilde{K}_{\rm odd}^{12} &=  \left[ \lambda_{2,  33} e^{i \beta_{2,  33} } \left( - i \tilde{t}_{12} \tilde{t}_{23}  \mathcal{S}_{yzx}^{\rm odd} + i \tilde{t}_{13} \mathcal{S}_{x0y}^{\rm odd} \right) \right. 
    \label{eq:8by8_K_odd_12} \\
    &\left. - \lambda_{2,  34} e^{i \beta_{2,  34}} \left( i \tilde{t}_{12} \tilde{t}_{23} \mathcal{S}_{yzy}^{\rm odd} + i \tilde{t}_{13} \mathcal{S}_{x0x}^{\rm odd} \right) \right] \lambda_{1,  12} e^{-i \beta_{1,  12} },  \nonumber  \\
    \tilde{K}_{\rm odd}^{21} &=  \left[ \lambda_{2,  33} e^{i \beta_{2,  33} } \left( -i \tilde{t}_{23}  \mathcal{S}_{0xx}^{\rm odd} + i \tilde{t}_{12} \tilde{t}_{13} \mathcal{S}_{zyy}^{\rm odd} \right) \right. \label{eq:8by8_K_odd_21} \\
    & \left.- \lambda_{2,  34} e^{i \beta_{2,  34}} \left( i \tilde{t}_{23}  \mathcal{S}_{0xy}^{\rm odd} + i \tilde{t}_{12} \tilde{t}_{13} \mathcal{S}_{zyx}^{\rm odd} \right) \right] \lambda_{1,  21} e^{-i \beta_{1,  21} },  \nonumber  \\
    \tilde{K}_{\rm odd}^{22} &=  \left[ \lambda_{2,  33} e^{i \beta_{2,  33} } \left( i \tilde{t}_{23}  \mathcal{S}_{0yx}^{\rm odd} + i \tilde{t}_{12} \tilde{t}_{13} \mathcal{S}_{zxy}^{\rm odd} \right)  \right. \label{eq:8by8_K_odd_22} \\
    & \left. + \lambda_{2,  34} e^{i \beta_{2,  34}} \left( i \tilde{t}_{23}  \mathcal{S}_{0yy}^{\rm odd} - i \tilde{t}_{12} \tilde{t}_{13} \mathcal{S}_{zxx}^{\rm odd} \right) \right] \lambda_{1,  22} e^{-i \beta_{1,  22} } , \nonumber 
\end{align}
where $\mathcal{S}_{abc}^{\rm odd}$ denotes the \rm odd-parity block in $\mathcal{S}_{abc}$. One can thus obtain the jump operator as  $\tilde{K}_{\rm odd} = \tilde{K}_{\rm odd}^{11} + \tilde{K}_{\rm odd}^{12} + \tilde{K}_{\rm odd}^{21} + \tilde{K}_{\rm odd}^{22}$.

To approach the structure of the jump operator in Eq.~\eqref{eq:8by8_jump_operator_f_g_form}, the tunneling phases are defined as
\begin{align}
& \beta _{1,  11} = \beta _{2,  33}+\frac{\pi }{2},  \beta _{1,  12} = \beta _{2,  33}+\pi ,  \label{eq:8by8_phases_setting_setup01} \\
& \beta _{1,  21} = \beta _{2,  33}-\frac{\pi }{2},  \beta _{1,  22} = \beta _{2,  33}+\pi ,  \beta _{2,  34} = \beta _{2,  33}+\frac{\pi }{2}. \nonumber 
\end{align}
The tunneling amplitudes between the Majorana boxes are set to $t_{12} = (t_{13}+t_{23})/(t_{13}-t_{23})$ for $t_{13} > t_{23}$. Moreover, defining
\begin{align}
    & \lambda_{1,  1 \pm} = \lambda_{1,  12} \pm \lambda_{1,  11}, \label{eq:8by8_notation_lambda_pm} \\
    & \lambda_{1,  2 \pm} = \lambda_{1,  21} \pm \lambda_{1,  22}, \nonumber \\
    & \lambda_{2,  3 \pm} = \lambda_{2,  33} \pm \lambda_{2,  34}, \nonumber
\end{align}
the jump operator becomes 
\begin{align}
    & \tilde{K}_{\rm odd} = \label{eq:8by8_jump_operator_phases_MBTunnel_determined} \\
    & \frac{t_{13}^2+t_{23}^2}{t_{23}-t_{13}} \left(
	\begin{array}{cccc}
		0 & \lambda_{1,  2-} \lambda_{2,  3-} & -\lambda_{1,  1+} \lambda_{2,  3-} & 0 \\
		\lambda_{1,  2+} \lambda_{2,  3+} & 0 & 0 & \lambda_{1,  1+} \lambda_{2,  3+} \\
		\lambda_{1,  1-} \lambda_{2,  3+} & 0 & 0 & \lambda_{1,  2-} \lambda_{2,  3+} \\
		0 & -\lambda_{1,  1-} \lambda_{2,  3-} & \lambda_{1,  2+} \lambda_{2,  3-} & 0 \\
	\end{array}
	\right). \notag
\end{align} 
For the tunneling amplitude, the last key condition needed to attain the structure of the ideal jump operator \eqref{eq:8by8_jump_operator_f_g_form} is given by
\begin{align}
    \lambda_{1,  1+} \lambda_{1,  1-} = \lambda_{1,  2+} \lambda_{1,  2-} \label{eq:8by8_lambda_setting_setup01} 
    \Leftrightarrow
      \lambda _{1,  12}^2 - \lambda _{1,  11}^2 = \lambda _{1,  21}^2 - \lambda _{1,  22}^2. 
\end{align}
For simplicity, we introduce the following symbols
\begin{align}
   \tilde{f} &= \frac{ \lambda_{1,1+} }{ \lambda_{1,2-} } = \frac{ \lambda_{1,2+} }{ \lambda_{1,1-} } \notag \\
   \tilde{g} &= \frac{ \lambda_{1,2-} }{ \lambda_{1,1-} } = \frac{ \lambda_{1,1+} }{ \lambda_{1,2+} }. \label{eq:8by8_notation_tilde_f_g}
\end{align}
which allow the jump operator \eqref{eq:8by8_jump_operator_phases_MBTunnel_determined} to be written as
\begin{align}
    & \tilde{K}_{odd} = \Lambda \left(
	\begin{array}{cccc}
		0 &  \tilde{g} \lambda _{2,  3-} & - \tilde{f} \tilde{g} \lambda _{2,  3-} & 0 \\
		\tilde{f} \lambda _{2,  3+} & 0 & 0 & \tilde{f} \tilde{g} \lambda _{2,  3+} \\
		\lambda _{2,  3+} & 0 & 0 &  \tilde{g} \lambda _{2,  3+} \\
		0 & - \lambda _{2,  3-} & \tilde{f} \lambda _{2,  3-} & 0 \\
	\end{array}
	\right), \label{eq:8by8_jump_operator_ideal_setup01} 
\end{align}
with $\Lambda = \lambda_{1,  1-}(t_{13}^2+t_{23}^2 )/ ( t_{23} - t_{13} )$. If we define $\tilde{f}_{1} = \tilde{g} \lambda_{2,  3-}$, $\tilde{f}_{2} = -\lambda_{2,  3-}$, $\tilde{g}_{1} = \tilde{f} \lambda_{2,  3+}$  and $\tilde{g}_{2} = \lambda_{2,  3+}$, the jump operator $\tilde{K}_{\rm odd}$ \eqref{eq:8by8_jump_operator_ideal_setup01} becomes equivalent to the jump operator \eqref{eq:8by8_jump_operator_f_g_form}. As anticipated, the dark sates of $\tilde{K}_{\rm odd}$, $\vert \tilde{ \psi }_{1} \rangle$ and $\vert \tilde{ \psi_{2} } \rangle$, have the form of $\vert \psi_{1} \rangle$ \eqref{eq:8by8_darkstate_psi1_f_g_form} and $\vert \psi_{2} \rangle$ \eqref{eq:8by8_darkstate_psi2_f_g_form},
\begin{align}
    \ket{ \tilde{ \psi_{1} } } &= \tilde{ N }_1 \left( -\tilde{g} \ket{001} + \ket{111} \right) \label{eq:8by8_dark_states_psi1_setup01} \\
    &=  \tilde{ N }_1 \left( \frac{ \lambda _{1,  22}-\lambda _{1,  21} }{ \lambda _{1,  12}-\lambda _{1,  11} } \ket{001} + \ket{111} \right), \nonumber \\
     \ket{ \tilde{ \psi_{2} } } &=  \tilde{ N }_2 \left( \tilde{f} \ket{010} + \ket{100} \right) \label{eq:8by8_dark_states_psi2_setup01} \\
     &= \tilde{ N }_2 \left( \frac{ \lambda _{1,  21} + \lambda _{1,  22} }{ \lambda _{1,  12}-\lambda _{1,  11} } \ket{010} + \ket{100} \right), \nonumber
\end{align}
From Eqs.~\eqref{eq:psi_3}-\eqref{eq:psi_4}, one can find that the other two eigenstates $ \vert \psi_{\pm} \rangle $ of $\tilde{K}_{\rm odd}$ \eqref{eq:8by8_jump_operator_ideal_setup01} merge and become a null vector in this case. 

Our choice of tunneling amplitudes ensures that the jump operator yields two parameter-dependent dark states within the odd-parity subspace. However, the tunneling system must avoid two scenarios: $\lambda_{2,  3 +} = 0$ and $\lambda_{2,  3 -} = 0$, i.e., the tunneling amplitudes with QD $2$ cannot be in phase, $\lambda_{2,  33} = \lambda_{2,  34}$, or exactly out of phase, $\lambda_{2,  33} = -\lambda_{2,  34}$. Since the dark space only contains one parameter-dependent dark state when $\lambda_{2,  3\pm} = 0$, this scenario cannot support a braiding transformation. Therefore, we adiabatically vary the tunneling amplitudes $\lambda_{1,  2\pm}$ while keeping $\lambda_{2,  3\pm} $ as a non-vanishing constant during braiding.

\subsection{The braiding transformation} \label{subsec:threeboxes_briding_manipulation_setup01}

The parameters $\lambda_{1,  2\pm}$ are changed adiabatically, and $\lambda_{1,  1\pm}$ changes accordingly due to Eq.~\eqref{eq:8by8_lambda_setting_setup01}. With the time-periodic boundary condition from time $t_{0}$ to $t_{f}$, we establish the following polar coordinate systems for the tunnel amplitudes.
\begin{align}
    & \lambda _{1,  2-}(t)  = r_{2} \sin[ \theta_{2}(t) ],  \lambda _{1,  2+} (t) = r_{2} \cos[ \theta_{2}(t) ]; \label{eq:8by8_setup01_lambda2_polar} \\
    & \lambda _{1,  1-} (t) = r_{1} \cos[ \theta_{1}(t) ],  \lambda _{1,  1+} (t) = r_{1} \sin[ \theta_{1}(t) ], \label{eq:8by8_setup01_lambda1_polar}
\end{align}
where $\theta_{l} (t_{f}) = \theta_{l} (t_{0}) + 2 \pi$ for $l \in \{1,  2\}$. The parameters should follow $r_{2}^{2} \cos( \theta_{2} )\sin( \theta_{2} ) \equiv r_{1}^{2} \cos( \theta_{1} )\sin( \theta_{1} )$ due to the requirement~\eqref{eq:8by8_lambda_setting_setup01}.

As shown in Sec.~\ref{subsec:stabilization_in_three_MBs}, the pure steady state $\tilde {\rho}_{s}$ in this setup can be formulated as $\rho_{s}$ in Eq.~\eqref{DM:8by8_steady_state_pure}. Analogously to the function $F(f, g)$ \eqref{eq:8by8_case_pi_F} in the steady state $\rho_{s}$, the function $\tilde{F}$ in this case is then translated into
\begin{align}
    \tilde{F} (\lambda_{1,  21},  \lambda_{1,  22}) = \sin \left[ \frac{1}{4} \sin^{-1} \left( \frac{\lambda _{1,  21}-\lambda _{1,  22}}{\sqrt{ 2 \lambda _{1,  21}^2 + 2 \lambda _{1,  22}^2} } \right) \right]. \label{eq:8by8_setup01_F_lambda}
\end{align}
Using Eq.~\eqref{eq:8by8_setup01_lambda2_polar}-\eqref{eq:8by8_setup01_lambda1_polar}, one can express $\tilde{F}$ as a function of $\theta_2$ as $\tilde{F} (\theta_{2}) = \sin( \theta_{2} /4)$. Referring to the definition of the Bloch vector in Eq.~\eqref{BV:8by8_bloch_vector_in_terms_of_F}, this results in $ \tilde{ \bm{n} } = ( \sin[\theta_{2}/2],  0,  \cos[\theta_{2}/2] )$. As anticipated, the Bloch vector is parametrized by $\theta_{2}$, indicating its dependence on $\theta_{2}$ rather than time $t$. 
The steady state with this Bloch vector reaches $\vert \tilde{\psi}_{2} \rangle$ from $\vert \tilde{\psi}_{1} \rangle$ after the parameters have been varied along one cycle. This process thus gives rise to the braiding operator $B = \sigma_{y}$ as described in Eq.~\eqref{eq:8by8_braiding_operator_sigmay}.

\begin{figure*}[ht]
	\centering
	\includegraphics[width=0.4\textwidth]{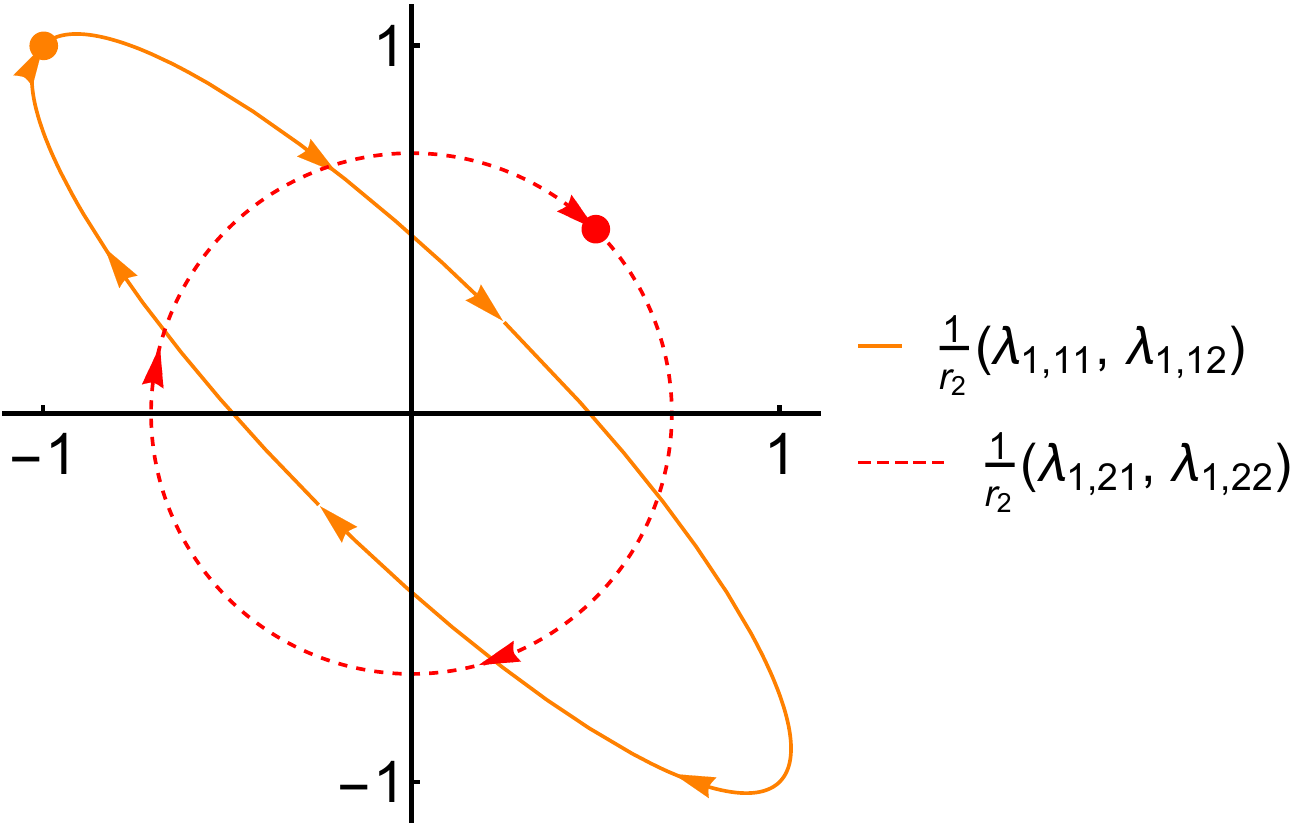}
    \includegraphics[width=0.35\textwidth]{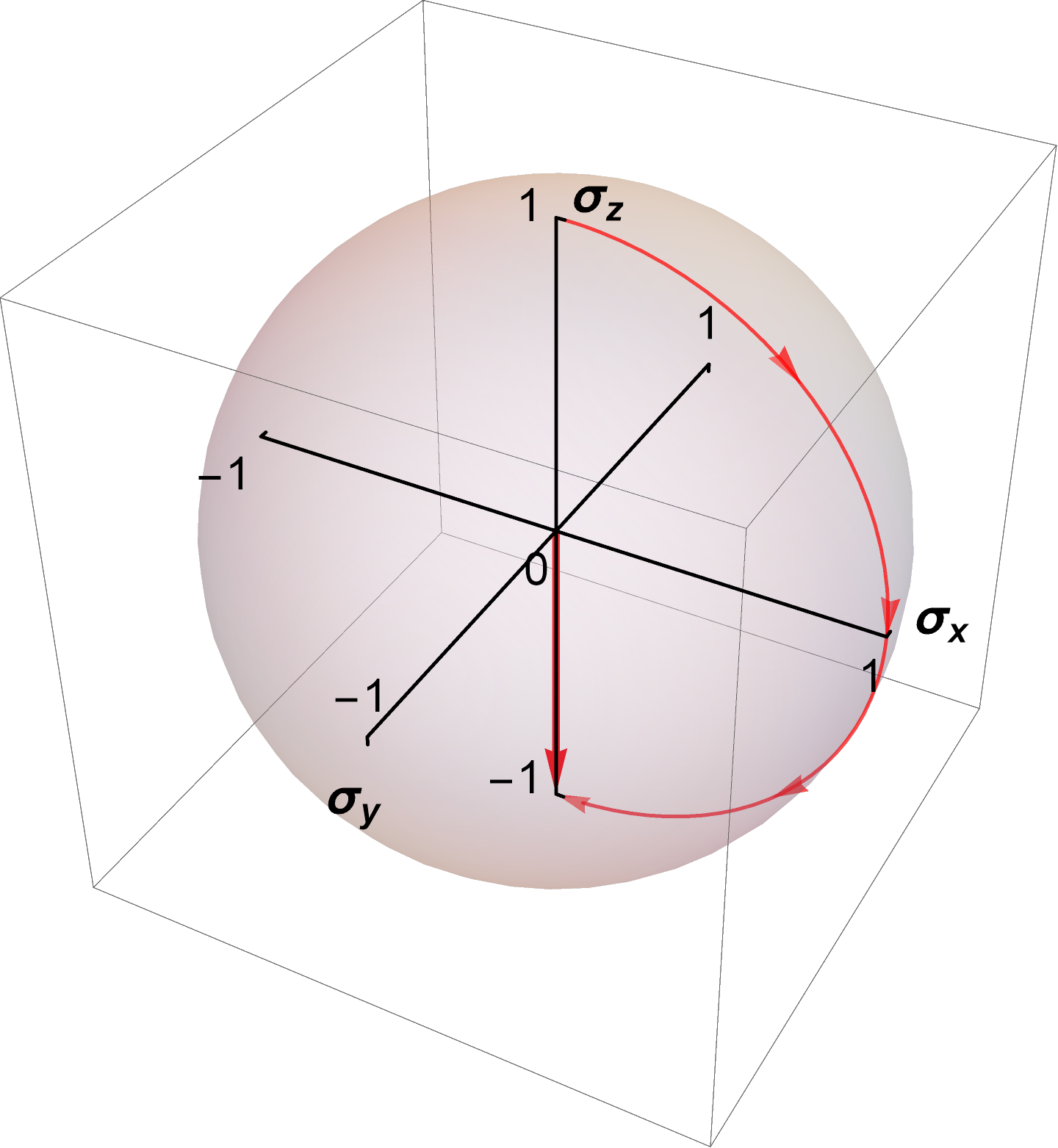}
	\caption{The left diagram shows how the tunneling amplitudes vary from $\theta_{2}(t_{0}) = 0$ to $\theta_{2}(t_{f}) =2 \pi$ when $r_{1} = 2 r_{2}$ and $\theta_{1}(t) = \theta_{2}(t)$ in Eqs.~\eqref{eq:8by8_setup01_lambda2_polar}-\eqref{eq:8by8_setup01_lambda1_polar}. The solid dots in the cyclic adiabatic evolution represent the initial and final values of the parameters. The Bloch vector $\tilde{ \bm{n} } = ( \sin[\theta_{2}/2],  0,  \cos[\theta_{2}/2] )$ undergoes the rotation depicted in the right figure.}
	\label{figs:parameters_and_BlochVector_vary_in_braiding_protocol}
\end{figure*}

In Fig.~\ref{figs:parameters_and_BlochVector_vary_in_braiding_protocol}, we present a numerical example where $r_{1} = 2 r_{2}$ and $\theta_{1}(t) = \theta_{2}(t)$, with $\theta$ linearly varying with time. This figure shows the trajectory of the Bloch vector alongside the periodic change in parameters.

To summarize, in this section, we first ascertained that braiding requires a minimum of three Majorana boxes, which collectively form an open quantum system potentially possessing two parameter-dependent dark states with the same parity. 
In Sec.~\ref{subsec:three_MBs_ideal_jump_operator}, we introduced the ideal jump operator \eqref{eq:8by8_jump_operator_f_g_form}, which encompasses two dark states \eqref{eq:8by8_darkstate_psi1_f_g_form} - \eqref{eq:8by8_darkstate_psi2_f_g_form} within the odd-parity subspace. Following the Lindblad equation with this jump operator, the dissipation should lead the reduced system into a state within the subspace spanned exclusively by these two dark states, as depicted in Eqs.~\eqref{DM:8by8_pure_state_F_alpha12} and \eqref{DM:8by8_steady_state_pure}. 
To preserve the chiral symmetry in the steady state, the function 
$F$ in Eq.~\eqref{DM:8by8_steady_state_pure} must be real and must obey
$F \in [-1,\; 1]$. Due to the presence of chiral symmetry, topological order emerges in the steady state, characterized by the Bloch vector \eqref{BV:8by8_bloch_vector_in_terms_of_F} described in Sec.~\ref{subsec:stabilization_in_three_MBs}.

Based on the ideal jump operator \eqref{eq:8by8_jump_operator_f_g_form}, we constructed the tunneling system depicted in Fig.~\ref{fig:8by8_setup01}, which represents the simplest setup leading to the desired structure of the ideal jump operator. By an appropriate choice of tunneling parameters, the jump operator \eqref{eq:8by8_jump_operator_ideal_setup01} in the setup of Fig.~\ref{fig:8by8_setup01} has two parameter-dependent and odd-parity dark states as shown in Eqs.~\eqref{eq:8by8_dark_states_psi1_setup01}-\eqref{eq:8by8_dark_states_psi2_setup01}. However, one should keep in mind to avoid in-phase and opposite-phase tunneling amplitudes between QD $2$ and MBSs, i.e., $\lambda_{2, 33} \neq \pm \lambda_{2, 34}$ because in either of these cases braiding is not feasible.

\section{Conclusion} \label{sec:conclusion_and_outlooks}

We have considered a driven-dissipative system consisting of Majorana box quits. In such a system, the Majorana bound states are coupled via electron tunneling to quantum dots, which are in turn driven by an applied AC voltage. Dissipation is brought about by the coupling to the electromagnetic environment. It had been shown before that this combination of drive and dissipation can stabilize certain Majorana qubit states as dark states of the corresponding Lindblad equation. 

We have considered the question of whether topological transformations, i.e., braiding, can be executed within such a dark-state subspace. In a non-Abelian exchange due to braiding, rotations between different, degenerate qubit states are achieved due to an adiabatic change of system parameters. Moreover, the result of the braiding transformation should be topological, in the sense that it depends only on the topological winding number of the parameter path. 

We showed that such a braiding transformation is indeed possible in a driven-dissipative Majorana box system consisting of three Majorana boxes. The latter contains a four-dimensional odd-parity subspace. Moreover, for a suitable choice of tunneling couplings, such a system can host a two-dimensional dark-state subspace in which a braiding transformation can be implemented. To achieve such braiding, certain parameters of the system have to be tuned to special symmetric values. However, the remaining free parameters still allow an adiabatic change of parameters along closed loops in parameter space, and the resulting braiding transformation is topological in this reduced parameter space.

\section{Acknowledgements}

The authors would like to acknowledge useful discussions with Johan Ekström. SSK acknowledges financial support from the National Research Fund Luxembourg under grant C18/MS/12704391/QUTHERM.


\appendix

\section{The solid angle} \label{app:solid_angle}

When the parameters are driven by time-periodic functions from $t=0$ to $t=T$, one obtains the solid angle given in Eq.~\eqref{SA:2by2_general_xz_plane}. The Bloch vector \eqref{BV:2by2_pure_state} has unit length and is orthogonal to the vector $\bm{a} = (0, 1, 0)$. The solid angle can then be calculated as follows
\begin{align}
\Omega &= \int_{0}^{T} dt \; \left[ \bm{n} (k_x, k_z) \times \partial_t \bm{n}(k_x, k_z) \right] \cdot \bm{a} \notag \\
&= \int_{0}^{T} dt \left( n_z \partial_t n_x - n_x \partial_t  n_z \right).
\end{align}
Due to $n_{x}^2 + n_{z}^2 = 1$, we can introduce polar coordinates, $n_{x}(t) = \sin \Theta(t)$ and $n_{z} (t) = \cos \Theta (t)$. Then, we find
\begin{align}
    \partial_t \ln ( e^{ i \Theta } ) 
&=
    \partial_t \ln (n_{z} + i n_{x}) = \frac{1}{ n_{z} + i n_{x} } \partial_t \left( n_{z} + i n_{x} \right) \notag \\
&= \frac{ n_{z} - i n_{x} }{ n_{z}^2 + n_{x}^2 } \partial_t \left( n_{z} + i n_{x} \right) \nonumber \\
&= i \left( n_{z} \partial_t  n_{x} -  n_{x} \partial_t n_{z} \right).
\end{align}
Using this in the integral of the solid angle, we have
\begin{align}
\Omega &= -i \int_{0}^{T} dt \; \partial_t \ln \left( e^{i \Theta} \right) =  \int_{0}^{T} dt \; \partial_t \Theta(t), \nonumber \\
&= \Theta(T) - \Theta(0),
\end{align}
where $\Theta(t) = \arctan (n_{x} / n_{z}) = -\arctan (k_{z} / k_{x})$. 

\section{Stabilization of three entangled Majorana boxes in the odd-parity subspace} \label{app:three_boxes_Lindbladian_stabilization}

As shown in Sec.~\ref{subsec:three_MBs_ideal_jump_operator}, we only need to focus on the dissipation in the odd-parity sector, which corresponds to the manifold of the four odd-parity states $\ket{001}$, $\ket{010}$, $\ket{100}$ and $\ket{111}$. Without loss of generality, we can express an arbitrary $4 \times 4$ density matrix as follows,
\begin{align}
    \rho &= \sum_{n = 0} ^{3} a_{nn} \ket{n}\bra{n} + \sum_{n> m} \left[ \left( a_{nm} + i \; b_{nm} \right) \ket{n} \bra{m} + \text{h.c.}\right], \label{appeq:8by8_arbitrary_DM}
\end{align}
where $a_{nm}$ and $b_{nm}$ are real numbers and where $\{\ket{0}, \ket{1}, \ket{2}, \ket{3}\}$ are shorthands for the four odd-parity states. The matrix coefficients should satisfy $\sum_{n} a_{nn} \equiv 1$ and $ a_{nn} \geq 0$ for $n \in \{0,1,2,3\}$.

Using the jump operator~\eqref{eq:8by8_jump_operator_f_g_form}, the Lindblad operator $\mathcal{L}[K_{\rm odd}]$ drives the reduced system to the steady state $\rho_s$ when $\mathcal{L}[K_{\rm odd}] \rho_{s} = 0$. This condition dictates that the elements of the steady state density matrix should obey, in the case of the diagonal elements,
\begin{align}
a_{00} &= -g a_{03}, &  a_{11}  &= \frac{f^2 \left( a_{03} g^2+ a_{03}+g\right)}{\left(f^2+1\right) g}, \label{appeq:8by8_steady_state_akk} \\
a_{33} &= -\frac{ a_{03} }{g} & a_{22} &= \frac{ a_{03} g^2+ a_{03}+g}{f^2 g+g} \notag
\end{align}
and for the off-diagonal elements,
\begin{align}
a_{01} &= -f g a_{23}, & & a_{02}  = -g a_{23}, \label{appeq:8by8_steady_state_ajk} \\
a_{13} &= f a_{23}, & & \frac{ 1 + f^2 }{f} a_{12}  + \frac{ 1 + g^2 }{ -g } a_{03} = 1; \nonumber \\
b_{01} &= f g b_{23}, & & b_{02} = g b_{23} \label{appeq:8by8_steady_state_bjk} \\
b_{03} &= b_{12} = 0, & & b_{13} = f b_{23}. \nonumber 
\end{align}
We can define a function $F(f,g) \in [-1, 1]$ to replace $a_{12}$ and $a_{03}$ in the steady state as the following expressions.
\begin{align}
    a_{12} = F^2 \frac{f}{ 1+f^2 } \; \text{and} \; a_{03} = \left( 1 - F^2 \right) \frac{-g}{ 1+g^2 }, \label{appeq:8by8_steady_state_a03_a12}
\end{align}
such that it satisfies the last equation in Eq.~\eqref{appeq:8by8_steady_state_ajk}.
Then, one finds the following general form of the steady state density matrix in the odd-parity subspace,
\begin{widetext}
\begin{align}
\rho_{s} = \left(
\begin{array}{cccc}
	\frac{ \left( 1- F^2 \right) g^2 }{g^2+1} & f g (-a_{23} +i  b_{23} ) & g ( -a_{23} + i  b_{23} ) & \frac{\left(F^2-1\right) g}{g^2+1} \\
	f g ( -a_{23} - i  b_{23} ) & \frac{f^2 F^2}{f^2+1} & \frac{f F^2}{f^2+1} & f ( a_{23} +i  b_{23} ) \\
	g ( -a_{23} - i  b_{23} ) & \frac{f F^2}{f^2+1} & \frac{F^2}{f^2+1} &  a_{23} +i  b_{23}  \\
	\frac{\left(F^2-1\right) g}{g^2+1} & f ( a_{23} -i  b_{23} ) &  a_{23} -i  b_{23}  & \frac{1-F^2}{g^2+1} \\
\end{array}
\right). \label{appeq:8by8_general_steady_state_pure_and_mixed}
\end{align}
\end{widetext}

This dissipation should stabilize the two dark states in Eqs.~\eqref{eq:8by8_darkstate_psi1_f_g_form} and \eqref{eq:8by8_darkstate_psi2_f_g_form}. We thus change the basis of Eq.~\eqref{appeq:8by8_general_steady_state_pure_and_mixed} using, 
\begin{align}
	U &= \left( \ket{\psi_1} \; \ket{\psi_2} \; \ket{\psi_{3}} \; \ket{\psi_{4}} \right) \\
        &= \begin{pmatrix}
		-g /\sqrt{1+g^2} & 0 & 1/ \sqrt{1+g^2} & 0 \\
		0 & f / \sqrt{1+f^2} & 0 & -1/ \sqrt{1+f^2} \\
		0 & 1/ \sqrt{1+f^2} & 0 & f / \sqrt{1+f^2} \\
		1 / \sqrt{1+g^2} & 0 & g / \sqrt{1+g^2} & 0 \\
	\end{pmatrix}, \nonumber
\end{align}
where $ \vert \psi_{1} \rangle$ and $ \vert \psi_{2} \rangle$ are the dark states \eqref{eq:8by8_darkstate_psi1_f_g_form}-\eqref{eq:8by8_darkstate_psi2_f_g_form}, and $ \vert \psi_{3} \rangle$ \eqref{eq:psi_3} and $ \vert \psi_{4} \rangle$ \eqref{eq:psi_4} satisfy $\rho_{s} \vert \psi_{3} \rangle = \rho_{s} \vert \psi_{4} \rangle = 0$. The vectors $\vert \psi_{3} \rangle$ and $\vert \psi_{4} \rangle$ are found by diagonalizing $\rho_{s}$ \eqref{appeq:8by8_general_steady_state_pure_and_mixed}. As anticipated, $\rho_{s}$ is a semidefinite matrix with two zero eigenvalues, because the dark space consists of only two dark states. In this basis, all components of the steady state density matrix should vanish except those in the subspace of $\{ \vert \psi_{1} \rangle, \; \vert \psi_{2} \rangle \}$. Therefore,
\begin{align}
	\rho_{s} &= \left( 1-F^2 \right) \vert \psi_{1} \rangle \langle \psi_{1} \vert \label{appeq:8by8_steady_in_basis_psi_j} 
        +  r_{23} e^{-i \theta_{23}} \sqrt{(f^2+1) (g^2+1)} \vert \psi_{1} \rangle \langle \psi_{2} \vert \nonumber \\
    &+ 
        r_{23} e^{i \theta_{23}} \sqrt{(f^2+1) (g^2+1)} \vert \psi_{2} \rangle \langle \psi_{1} \vert
        + F^2 \vert \psi_{2} \rangle \langle \psi_{2} \vert, \nonumber
\end{align}
where $r_{23} = \text{sgn}(a_{23})\sqrt{a_{23} + b_{23}}$ and $\theta_{23} = \arctan(b_{23}/a_{23})$ with $\theta_{23} \in [0, \; \pi)$. To get a pure state, we can define $r_{23}$ as follows,
\begin{align}
r_{23} = \frac{ F\sqrt{ 1-F^2 } }{ \sqrt{(f^2+1) (g^2+1)} }.
\end{align}
As expected, the pure state becomes the combination of the two dark states $\vert \psi_{1} \rangle$ and $\vert \psi_{2} \rangle$, 
\begin{align}
    \rho_{s} = \left( \sqrt{1-F^{2}} e^{i\alpha_{1}} \vert \psi_{1} \rangle +   F e^{i\alpha_{2}} \vert \psi_{2} \rangle \right) 
    \otimes \text{h.c.} \label{appeq:8by8_pure_state_F_alpha12}
\end{align}
where $\alpha_{1} - \alpha_{2} = \theta_{23}$, and we assumed that $F$ and $r_{23}$ have the same sign.

To ensure a pure state with chiral symmetry, one possibility is to choose $\theta_{23} = 0$, or $b_{23} = 0$ and $a_{23}^2 = ( F^2 - F^4 ) / [ (f^2+1) (g^2+1) ]$. This can be done by changing the relative phase between the two dark states $\vert \psi_{1} \rangle$ and $\vert \psi_{2} \rangle$. Then, the steady state~\eqref{appeq:8by8_steady_in_basis_psi_j} becomes
\begin{align}
	\rho_{s} &= \left( 1-F^2 \right) \vert \psi_{1} \rangle \langle \psi_{1}|  + F \sqrt{1 - F^2} \; \vert \psi_{1} \rangle     \langle \psi_{2} \vert \label{appeq:8by8_pure_state_expressed_by_F}\\
    & + F \sqrt{1 - F^2} \; \vert \psi_{2} \rangle \langle \psi_{1} \vert + F^2 \vert \psi_{2} \rangle \langle \psi_{2} \vert, \nonumber
\end{align}
whose Bloch vector $( 2  F \sqrt{ 1 - F^2 },  0, 1 - 2 F^2 )$ rotates in the $\sigma_{x}-\sigma_{z}$ plane. 

\section{The jump operator in the tunneling system of three entangled Majorana boxes} \label{app:three_MBs_jump_operator_setup01}

To determine the corresponding jump operators in a system with three Majorana boxes, we generalize the derivation presented in Sec.~\ref{subsec:dissipation_in_Hamiltonian}, i.e., we apply a SW transformation to the tunneling Hamiltonian to obtain an effective cotunneling Hamiltonian. Subsequently, we can extract the operator in the Majorana sector which becomes the jump operator after tracing over the other degrees of freedom within a Born-Markov approximation.
	
In the case of three Majorana boxes, the Hamiltonian of the open system is given by
\begin{align}
    H(t) = H_{\rm box} + H_{\rm QD} + H_{\rm env} + H_{\rm drive}(t) + H_{\rm tun, e} + H_{\rm tun, \gamma},
\end{align}
where $H_{\rm box}$, $H_{\rm QD}$, $H_{\rm env}$, and $H_{\rm drive}(t)$ represent the Hamiltonian of the Majorana box, quantum dots, environment, and the time-dependent driving field, respectively. For three Majorana boxes, $H_{\rm box} = E_{C} \sum_{n=1}^{3} (N_{n} - N_{g})^2$. The tunneling Hamiltonian consists of two contributions. Electron tunneling between the quantum dots and Majorana box $n$ is represented by,
\begin{align}
H_{\rm tun,e} &= t_{0} \sum_{n=1}^3 \sum_{ j=1}^2 \sum_{\mu=1}^4 \lambda_{j, n \mu} e^{ -i \beta_{j, n \mu} } e^{-i\hat{\phi}_{n} + i\delta_{j} + i\epsilon_{j} t} d_{j}^{\dagger} \gamma_{n \mu}  \label{appeq:8by8_tunneling_Hamiltonian} + \text{h.c.},
\end{align}
where direct tunneling between the two Majorana boxes (the box $n$ and the box $n'$) is described by,
\begin{align}
H_{\rm tun,\gamma} &= E_{C} \sum_{n,n' = 1}^3 \sum_{\mu, \nu =1}^4 i \tilde{t}_{n n'} \gamma_{n \mu} \gamma_{n' \nu}, \label{appeq:8by8_tunnels_between_MBs} 
\end{align}
where the dimensionless parameter $\tilde{t}_{n n'} = t_{n n'} / E_{C}$ and $t_{nn'}$ is the tunnel coupling energy between the boxes $n$ and $n'$. Here, we assume that $t_{nn} = 0$ and $t_{n n'} > 0$ for all amplitudes.

To perform the SW transformation, we once again assume the energy scales of tunnel couplings, $t_{0}$ and $t_{nn'}$, to be small. Applying the rotating wave approximation to remove fast oscillating terms in the driving Hamiltonian, the unperturbed Hamiltonian becomes time-independent and can be expressed as follows,
\begin{align}
H_{0} &= H_{\rm box} + H_{\rm QD} + H_{\rm env} + H_{\rm drive} \notag \\
&= E_{C} \sum_{n} \left( N_{n} -N_{g} \right)^2 + \sum_{j =1,2} \epsilon_{j} d_{j}^{\dagger} d_{j} + \sum_{m} E_{m} b_{m}^{\dagger} b_{m} \label{appeq:H0_SW_transformation} \\
& + \mathcal{A} (d_{2}^{\dagger} d_{1} + h.c.), \notag
\end{align}
where $\mathcal{A}$ is the amplitude of the driving field. 

We perform an SW transformation to the effective Hamiltonian $H_{\rm eff} = e^{S} (H_0 + H_{\rm tun}) e^{-S}$ where $H_{\rm tun} = H_{\rm tun, e} + H_{\rm tun, \gamma}$. For three Majorana boxes, it is necessary to go up to the fourth order in the expansion,
\begin{align}
H_{\rm eff} &\approx H_{0} + H_{\rm tun} + \left[ S, H_{0}\right] + \left[S, H_{\rm tun} \right] \label{appeq:SW_transformation_Taylor_expansion_upto_fourth_order} \\
&+ \frac{ 1 }{ 2 }\left[S, \; \left[ S, H_{0} \right]\right] + \frac{ 1 }{ 2 }\left[S, \left[ S, H_{\rm tun} \right]\right] \notag \\
&+ \frac{ 1 }{ 6 }\left[S, \; \left[ S, \left[ S, H_{0} \right] \right]\right] + \frac{ 1 }{ 6 }\left[S, \left[ S, \left[ S, H_{\rm tun} \right] \right]\right] \notag \\
& + \frac{ 1 }{ 24 }\left[S, \; \left[ S, \left[ S, \left[ S, H_{0} \right] \right] \right]\right]. \nonumber
\end{align}
In the SW transformation, the Hamiltonian can be diagonalized to first order in the perturbation by finding a generator $S$ which satisfies,
\begin{align}
H_{\rm tun} = - \left[ S, H_{0} \right]. \label{appeq:S_generator_SW_transformation}
\end{align} 
With this generator the expansion becomes
\begin{align}
H_{\rm eff} &\approx
H_{0} + \frac{ 1 }{ 2 } \left[S, \; H_{\rm tun} \right] + \frac{ 1 }{ 3 }\left[S, \; \left[ S, \; H_{\rm tun} \right]\right] \label{appeq:SW_transformation_Taylor_expansion_upto_fourth_order_simplified} \\
&+ \frac{ 1 }{ 8 }\left[S, \; \left[ S, \; \left[ S, \; H_{\rm tun} \right] \right]\right]. \notag
\end{align}
To find the generator $S$ satisfying Eq.~\eqref{appeq:S_generator_SW_transformation} explicitly, it is convenient to diagonalize the Hamiltonian $H_{0}$ \eqref{appeq:H0_SW_transformation}. If we write $H_{0}$ in a block-diagonal form, we only need to diagonalize it in the subspace of the quantum dot operators because $H_{\rm box}$ and $H_{\rm env}$ are already diagonal matrices. One finds
\begin{align}
    H_{\rm QD} + H_{\rm drive} = \begin{pmatrix}
        \epsilon_{1} & 0 & 0 & \mathcal{A} \\
        0 & -\epsilon_{1} & -\mathcal{A} & 0 \\
        0 & -\mathcal{A} & -\epsilon_{2} & 0 \\
        \mathcal{A} & 0 & 0 & \epsilon_{2} 
      \end{pmatrix} = U_{0} H_{d} U_{0}^{-1} ,
\end{align}
where $H_d = \text{diag}(- \eta_{+},\eta_{-}, -\eta_{-}, \eta_{+} )$ and 
\begin{widetext}
\begin{align}
U_{0}& = \begin{pmatrix}
		0 & \frac{\eta_{+} - 2 \epsilon_{2}}{\sqrt{(\eta_{+} - 2 \epsilon_{2})^{2} + 4 \mathcal{A}^{2} } }  & \frac{ 2 \mathcal{A} }{\sqrt{(\eta_{+} - 2 \epsilon_{2})^{2} + 4 \mathcal{A}^{2} } }  & 0  \\
		\frac{\eta_{-} - 2 \epsilon_{2}}{\sqrt{(\eta_{+} - 2 \epsilon_{1})^{2} + 4 \mathcal{A}^{2} } }  & 0 & 0 & \frac{ 2 \mathcal{A} }{\sqrt{(\eta_{+} - 2 \epsilon_{1})^{2} + 4 \mathcal{A}^{2} } } \\
			0 & \frac{ \eta_{-} - 2 \epsilon_{2} }{\sqrt{(\eta_{+} - 2 \epsilon_{1})^{2} + 4 \mathcal{A}^{2} } } & \frac{ 2 \mathcal{A} }{\sqrt{(\eta_{+} - 2 \epsilon_{1})^{2} + 4 \mathcal{A}^{2} } } & 0 \\
			\frac{ \eta_{+} - 2 \epsilon_{2} }{\sqrt{(\eta_{+} - 2 \epsilon_{2})^{2} + 4 \mathcal{A}^{2} } } & 0 & 0 & \frac{ 2 \mathcal{A} }{\sqrt{(\eta_{+} - 2 \epsilon_{2})^{2} + 4 \mathcal{A}^{2} } }
		\end{pmatrix}.
\end{align}
\end{widetext}
The eigenvalues of the diagonal matrix are given by $\eta_{\pm} = (\epsilon_{1} + \epsilon_{2} \pm \sqrt{4 \mathcal{A}^{2} + (\epsilon_{1} - \epsilon_{2})^{2}})/2$. Subsequently, the total Hamiltonian is transformed into this new basis, and the effective Hamiltonian is found to be
\begin{align}
H_{\rm eff}' = e^{S} \left( H_{0}' + H_{\rm tun} ' \right) e^{-S},  
\end{align}
where $H_{0} ' = U_{0} H_{0} U_{0}^{-1}$ and $H'_{\rm tun} = U_{0} H_{\rm tun} U_{0}^{-1}$. For the diagonal matrix $H_{0}'$, the components of the commutator in Eq.~\eqref{appeq:S_generator_SW_transformation} can be written as
\begin{align}
\left( H'_{\rm tun} \right)_{jk} = - \left[S, H_{0}' \right]_{jk} = \left( \epsilon'_{j} - \epsilon_{k}' \right) S_{jk}, \label{appeq:S_H0prime_commutator}
\end{align}
where $\epsilon_{j}'$ and $\epsilon_{k}'$ are the diagonal elements in $H_{0}'$, which depend on $E_{C}$, $E_{m}$, and $\eta_{\pm}$ in the subspace of MBSs, bath, and quantum dots, respectively. Using this to determine $S$, one can perform the SW transformation in Eq.~\eqref{appeq:SW_transformation_Taylor_expansion_upto_fourth_order_simplified}. For the second-order term, one finds, as they are in the different basis.
\begin{align}
\left( H'^{(2)}_{\rm cot} \right)_{jk} &= \frac{1}{2} \left[S, \; H'_{\rm tun} \right]_{jk} \\
&= \frac{1}{2} \sum_{l} \left( H'_{\rm tun} \right)_{jl} \left( H'_{\rm tun} \right)_{lk} \left(  \frac{ 1 }{ \epsilon'_{j} - \epsilon'_{l} } + \frac{ 1 }{\epsilon'_{k} - \epsilon'_{l} } \right) .
\end{align}
The low-energy regime imposes the condition that no energy scale exceeds the charging energy $E_{C}$ and the superconductivity gap $\Delta$. Therefore, the second-order term can be approximated as
\begin{align}
\left( H'^{(2)}_{\rm cot} \right)_{jk} \approx 
\frac{1}{E_{C}} \left( U_{0} H_{\rm tun} H_{\rm tun} U_{0}^{-1} \right)_{jk}.
\end{align}
For the third-order term in the expansion \eqref{appeq:SW_transformation_Taylor_expansion_upto_fourth_order_simplified}, one can use the second-order term and finds
\begin{align}
&\left( H'^{(3)}_{\rm cot} \right)_{jk} = \frac{1}{3} \left[S, \; \left[S, \; H'_{\rm tun} \right] \right]_{jk}  = \frac{1}{3} \left[S, \; 2 H_{cot}^{(2)} \right]_{jk} \\
&= \frac{2}{3 E_{C}} \sum_{l} \left( H'_{\rm tun} \right)_{jl} \left( H'_{\rm tun} H'_{\rm tun} \right)_{lk} \left[ \frac{ 1 }{\epsilon'_{j} - \epsilon'_{l} }  + \frac{ 1 }{\epsilon'_{k} - \epsilon'_{l} } \right]
\end{align}
In the low-energy regime, one can again approximate the third term as
\begin{align}
\left( H'^{(3)}_{\rm cot} \right)_{jk} \approx \frac{4}{3 E_{C}^2}  \left( U_{0} H_{\rm tun} H_{\rm tun} H_{\rm tun} U_{0}^{-1} \right)_{jk} .
\end{align}
Using the same method leads to the fourth term as follows.
\begin{align}
\left( H'^{(4)}_{\rm cot} \right)_{jk} \approx \frac{1}{E_{C}^{3} }  \left[ U_{0} \left( H_{\rm tun} \right)^{4} U_{0}^{-1} \right]_{jk} .
\end{align}

In the case of three Majorana boxes, the cotunneling Hamiltonian is obtained in the basis before the diagonalization of $H_{0}$,
\begin{align}
H_{\rm cot} = \sum_{r=2}^{4} \left( \frac{1}{(r-1)!} - \frac{1}{r!} \right)\left( \frac{2}{ E_{C} } \right)^{r-1} \left( H_{\rm tun} \right)^{r}.
\end{align}
However, some terms in the above formula do not describe full trajectories between the two quantum dots. For instance, because $H_{\rm tun} = H_{\rm tun, e} + H_{\rm tun, \gamma}$, $H_{\rm cot}^{(2)}$ turns into
\begin{align}
\frac{1}{E_{C}} \left( H_{\rm tun} H_{\rm tun} \right) =& \frac{1}{E_{C}} \left( H_{\rm tun, e} H_{\rm tun, e} + H_{\rm tun, e} H_{\rm tun, \gamma} \right. \\
 & \left. + H_{\rm tun, \gamma} H_{\rm tun, e} + H_{\rm tun, \gamma} H_{\rm tun, \gamma} \right). \notag
\end{align}
Only the first term, $(H_{\rm tun, e} H_{\rm tun, e})$, describes a complete trajectory between the two dots, similar to the scenario of a single Majorana box. Terms containing incomplete trajectories would only lead to subleading corrections to higher-order tunneling processes. Therefore, retaining only this term to second order, we find the second-order cotunneling Hamiltonian,
\begin{align}
H_{\rm cot}^{(2)} \approx \frac{ t_{0}^2 }{E_{C}} \sum_{n} & \left( e^{ i \delta_{12} + i \epsilon_{21} } A_{12}^{n} d_{2}^{\dagger} d_{1} + \text{h.c.} \right). \label{appeq:QD_QD_cotunneling_single_box}
\end{align}
The operator $A_{jk}^{n}$ denotes the dynamics within the Majorana box $n$, where the transport occurs between quantum dots $j$ and $k$.
\begin{align}
& A_{jk}^{n} = \sum_{ \mu < \nu } \Lambda_{jk}^{n \mu,n \nu} \gamma_{n \mu} \gamma_{n \nu} \label{appeq:operator_A_jk_n} \\
& \Lambda_{jk}^{n \mu,n \nu} = \lambda_{j, n \mu} \lambda_{k, n \nu} e^{ i \left( \beta_{k, n \nu} - \beta_{j, n \mu} \right)} - \lambda_{j, n \nu} \lambda_{k, n \mu} e^{ i \left( \beta_{k, n \mu} - \beta_{j, n \nu} \right)}. \label{appeq:Lambda_tunneling_parameters}
\end{align}
The subscript $n \in \{ 1, \;, 2, \; 3\}$ indicates the box, and $j, \; k \in \{1, \; 2\}$ represent the quantum dots. The complex parameter $\Lambda_{jk}^{n \mu,n \nu}$ contains the tunnel couplings between quantum dots $j$, $k$ and MBSs $\gamma_{n \mu}$ and $\gamma_{n \nu}$.
	
For the third-order term, one can find the complete transport between two dots as
\begin{align}
H_{cot}^{(3)} &\approx \frac{4}{ 3 E_{C}^2 } H_{tun, e} H_{tun, \gamma} H_{tun, e} \\
&= \frac{4 t_{0}^2 }{3 E_{C} } \sum_{n \neq n'} \left( e^{ i \delta_{12} + i \epsilon_{21} } A_{12}^{n, n'} d_{2}^{\dagger} d_{1} + \text{h.c.} \right) , \label{appeq:QD_QD_cotunneling_two_boxes} \\
\text{where } & A_{jk}^{n, n'} = i \tilde{t}_{n,n'} \sum_{ \mu < \nu } \sum_{ \mu' < \nu' } \Lambda_{jk}^{n \mu, n' \nu'} \gamma_{n \mu} \gamma_{n \nu} \gamma_{n' \mu'} \gamma_{n' \nu'}. \label{appeq:operator_A_jk_nn'}
\end{align}
In the operator $A_{jk}^{n,n'}$ \eqref{appeq:operator_A_jk_nn'}, the trajectory starts from quantum dot $j$, goes to Majorana box $n$, then proceeds to Majorana box $n'$ through the tunnel between the MBSs $\gamma_{n \nu}$ and $\gamma_{n' \mu'}$, and finally reaches quantum dot $k$. For the fourth order, the complete transport is written as 
\begin{align}
H_{cot}^{(4)} \approx \frac{1}{ E_{C}^3 } H_{tun, e} & H_{tun, \gamma} H_{tun, \gamma} H_{tun, e} \\
= \frac{ t_{0}^2 }{ E_{C} } \sum_{n \neq n' \neq n''} &  \left( e^{ i \delta_{12} + i \epsilon_{21} } A_{12}^{n, n', n''} d_{2}^{\dagger} d_{1} + \text{h.c.} \right) \label{appeq:QD_QD_cotunneling_three_boxes} \\
\text{where } A_{jk}^{n, n', n''} =& - \tilde{t}_{n,n'} \tilde{t}_{n',n''} \sum_{ \mu < \nu } \sum_{ \mu' < \nu' } \sum_{ \mu'' < \nu'' } \Lambda_{jk}^{n \mu, n'' \nu''} \label{appeq:operator_A_jk_nn'n''} \\
& \times \gamma_{n \mu} \gamma_{n \nu} \gamma_{n' \mu'} \gamma_{n' \nu'} \gamma_{n'' \mu''} \gamma_{n'' \nu''}.  \notag
\end{align}
The trajectory in the fourth-order term involves three boxes $n$, $n'$, and $n''$. Combining these three perturbation terms in Eqs. \eqref{appeq:QD_QD_cotunneling_single_box}, \eqref{appeq:QD_QD_cotunneling_two_boxes}, and \eqref{appeq:QD_QD_cotunneling_three_boxes}, one can obtain the cotunneling Hamiltonian from the SW transformation.  Furthermore, it is observed that the operators $A_{jk}$ in Eqs. \eqref{appeq:operator_A_jk_n}, \eqref{appeq:operator_A_jk_nn'}, and \eqref{appeq:operator_A_jk_nn'n''} serve as the building blocks for the jump operator in the Lindblad equation of the Majorana section after applying the Born-Markov approximation.

In an open system like Fig. \ref{fig:8by8_setup01}, the following Majorana bilinears represent the corresponding Pauli operators of resulting in the Majorana box $n$.
\begin{align}
\chi_x^{n} =  i \gamma_{n2} \gamma_{n4}, \; \chi_y^{n} = i \gamma_{n3} \gamma_{n2}, \; \chi_z^{n} = i \gamma_{n4} \gamma_{n3}.
\end{align}
For simplicity in the subsequent calculation, we define a matrix $\mathcal{S}_{abc}$ as follows.
\begin{align}
    & \mathcal{S}_{abc} = \mathcal{T} \cdot \left( \chi_{a}^{1} \otimes \chi_{b}^{2} \otimes \chi_{c}^{3} \right) \cdot \mathcal{T}^{-1}, \\
    & \text{ for } a, \; b, \; c \in \left\{ 0, \; x, \; y, \; z\right\}, \nonumber
\end{align}
where $\mathcal{T}$ is the unitary matrix rearranging the blocks in matrices according to the parity.

For three entangled boxes, there exist $4^{3}$ distinct $\mathcal{S}_{abc}$ matrices. However, certain matrices within this set vanish in the single-parity subspace, necessitating the exclusion of these trajectories since braiding does not change the parity of states. Through algebraic manipulations, one can identify a total of $16$ associated trajectories, with their corresponding $\mathcal{S}_{abc}$ matrices encompassed in the set $S$ as outlined below.
\begin{align}
    & \mathcal{S}_{abc} =  \mathcal{T} \cdot \left( \chi_{a}^{1} \otimes \chi_{b}^{2} \otimes \chi_{c}^{3} \right) .  \cdot \mathcal{T}^{\dagger}, \text{ for } (a, b, c) \in s, \label{appeq:8by8_tensorproducts_three_PauliMatrices_correlated_transport} \\
    &  	\text{where } s = \big\{ (0, x, x), \; (0, x, y), \; (0, y, x), \; (0, y, y), \nonumber \\
    & \quad \quad \quad \quad  \quad (x, 0, x), \; (x, 0, y), \; (x, z, x), \; (x, z, y), \nonumber \\
    & \quad \quad \quad \quad  \quad   (y, 0, x), \; (y, 0, y), \; (y, z, x), \; (y, z, y), \nonumber \\
    & \quad \quad \quad \quad  \quad (z, x, x), \; (z, x, y), \; (z, y, x), \; (z, y, y) \big\}. \nonumber
\end{align}  
Utilizing these $16$ trajectories, the tunneling system has been constructed as Fig. \ref{fig:8by8_setup01} in Sec. \ref{subsec:three_MBs_tunneling_system_and_braiding_protocol}.

To calculate the jump operator in the tunneling system illustrated in Fig. \ref{fig:8by8_setup01},  our focus is on the odd-parity subspace of MBSs exclusively. Considering the four tunnels originating from QD $1$ and connecting to Majorana qubits, we partition the odd-parity tunneling matrix $\tilde{K}_{odd}$ into four contributions: $\tilde{K}_{odd}^{11}$, $\tilde{K}_{odd}^{12}$, $\tilde{K}_{odd}^{21}$ and $\tilde{K}_{odd}^{22}$, with each superscript corresponding to a specific Majorana bound state. For instance, the computation of $\tilde{K}_{odd}^{11}$ can be visualized using the following tree diagram.
\begin{align}
	\gamma_{11} \begin{array}{l@{\ }l}
		\raisebox{-2.5ex}{$\nearrow$} & \gamma_{13} \xrightarrow{ t_{12} } \gamma_{24} \gamma_{23} \xrightarrow{ t_{23} } \gamma_{31} 
		\begin{array}{l@{\ }l}
			\raisebox{-0.7ex}{$\nearrow$} & \gamma_{33}  
			\vphantom{\dfrac1a}\\
			\raisebox{0.7ex}{$\searrow$}  & \gamma_{34}
		\end{array} 	\\
		\raisebox{2.5ex}{$\searrow$}  & \gamma_{14} \xrightarrow{ t_{13} } \gamma_{32}
		\begin{array}{l@{\ }l}
			\raisebox{-0.7ex}{$\nearrow$} & \gamma_{33}
			\vphantom{\dfrac1a}\\
			\raisebox{0.7ex}{$\searrow$}  & \gamma_{34}
		\end{array}
	\end{array}
 \label{appeq:8by8_setup01_lambda111_transport}				
\end{align}
From the operators $A_{jk}^{n,n'}$ \eqref{appeq:operator_A_jk_nn'} and $A_{jk}^{n,n',n''}$ \eqref{appeq:operator_A_jk_nn'n''}, one can express $\tilde{K}_{\rm odd}^{11}$ as 
\begin{align}
    \tilde{K}_{\rm odd}^{11} &= A_{12}^{1,3} + A_{12}^{1,2,3} \\
    \tilde{K}_{\rm odd}^{11} &= \left[ \lambda_{2, 33} e^{i \beta_{2, 33} } \left( i \tilde{t}_{12} \tilde{t}_{23}  \mathcal{S}_{xzx}^{\rm odd} + i \tilde{t}_{13} \mathcal{S}_{y0y}^{\rm odd} \right) \right. \label{appeq:8by8_K_odd_11} \\
     & \left. + \lambda_{2, 34} e^{i \beta_{2, 34}} \left( i \tilde{t}_{12} \tilde{t}_{23} \mathcal{S}_{xzy}^{\rm odd} - i \tilde{t}_{13} \mathcal{S}_{y0x}^{\rm odd} \right) \right] \lambda_{1, 11} e^{-i \beta_{1, 11} } , \nonumber
\end{align}
where $\mathcal{S}_{abc}^{\rm odd}$ denotes the odd-parity block in $\mathcal{S}_{abc}$. Applying the same method, the rest of the contributions in the jump operator $\tilde{K}_{\rm odd}$ are formulated as follows.
\begin{align}
    \tilde{K}_{\rm odd}^{12} &=  \left[ \lambda_{2, 33} e^{i \beta_{2, 33} } \left( - i \tilde{t}_{12} \tilde{t}_{23}  \mathcal{S}_{yzx}^{\rm odd} + i \tilde{t}_{13} \mathcal{S}_{x0y}^{\rm odd} \right) \right. 
    \label{appeq:8by8_K_odd_12} \\
    &\left. - \lambda_{2, 34} e^{i \beta_{2, 34}} \left( i \tilde{t}_{12} \tilde{t}_{23} \mathcal{S}_{yzy}^{\rm odd} + i \tilde{t}_{13} \mathcal{S}_{x0x}^{\rm odd} \right) \right] \lambda_{1, 12} e^{-i \beta_{1, 12} },  \nonumber  \\
    \tilde{K}_{\rm odd}^{21} &=  \left[ \lambda_{2, 33} e^{i \beta_{2, 33} } \left( -i \tilde{t}_{23}  \mathcal{S}_{0xx}^{\rm odd} + i \tilde{t}_{12} \tilde{t}_{13} \mathcal{S}_{zyy}^{\rm odd} \right) \right. \label{appeq:8by8_K_odd_21} \\
    & \left.- \lambda_{2, 34} e^{i \beta_{2, 34}} \left( i \tilde{t}_{23}  \mathcal{S}_{0xy}^{\rm odd} + i \tilde{t}_{12} \tilde{t}_{13} \mathcal{S}_{zyx}^{\rm odd} \right) \right] \lambda_{1, 21} e^{-i \beta_{1, 21} },  \nonumber  \\
    \tilde{K}_{\rm odd}^{22} &=  \left[ \lambda_{2, 33} e^{i \beta_{2, 33} } \left( i \tilde{t}_{23}  \mathcal{S}_{0yx}^{\rm odd} + i \tilde{t}_{12} \tilde{t}_{13} \mathcal{S}_{zxy}^{\rm odd} \right)  \right. \label{appeq:8by8_K_odd_22} \\
    & \left. + \lambda_{2, 34} e^{i \beta_{2, 34}} \left( i \tilde{t}_{23}  \mathcal{S}_{0yy}^{\rm odd} - i \tilde{t}_{12} \tilde{t}_{13} \mathcal{S}_{zxx}^{\rm odd} \right) \right] \lambda_{1, 22} e^{-i \beta_{1, 22} } . \nonumber 
\end{align}
By summing Eqs.~\eqref{appeq:8by8_K_odd_11}-\eqref{appeq:8by8_K_odd_22} up, we obtain the odd-parity block of the jump operator in the setup of Fig. \ref{fig:8by8_setup01}.
\begin{align}
    \tilde{K}_{\rm odd} = \tilde{K}_{\rm odd}^{11} + \tilde{K}_{\rm odd}^{12} + \tilde{K}_{\rm odd}^{21} + \tilde{K}_{\rm odd}^{22}.
\end{align}

\bibliography{topologydissipaton}

\end{document}